\documentclass[a4paper,fleqn]{cas-sc}

\usepackage[authoryear,longnamesfirst]{natbib}
\usepackage{lineno}
\usepackage{graphicx,subfigure,float}

\def\tsc#1{\csdef{#1}{\textsc{\lowercase{#1}}\xspace}}
\tsc{WGM}
\tsc{QE}

\begin{document}
\let\WriteBookmarks\relax
\def\floatpagepagefraction{1}
\def\textpagefraction{.001}

\shorttitle{}    

\shortauthors{}  

\title [mode = title]{Development and characterization of the JNE concentrator}  

\tnotemark[1] 

\tnotetext[1]{} 

%

\author[1]{Shuai Ouyang}






\affiliation[1]{organization={Key Laboratory of Particle Physics and Particle Irradiation (MOE), Institute of Frontier and Interdisciplinary Science, Shandong University},
            city={Qingdao},
            postcode={266237}, 
            state={Shandong},
            country={P.R. China}}

\author[2]{Yuzi Yang}

\cormark[1]


\ead{yangyz18@tsinghua.org.cn}



\affiliation[2]{organization={Department of Engineering Physics \& Center for High Energy Physics, Tsinghua University},
            city={Beijing},
            postcode={100084}, 
            country={P.R. China}}
\author[1]{Yang Zhang}
\cormark[1]


\ead{yangzhangsdu@email.sdu.edu.cn}



            
\author[2]{Aiqiang Zhang}
\author[3]{Haoyan Yang}
\author[2]{Changxu Wei}
\author[4]{Yuhao Liu}
\author[2]{Zhe Wang}
\author[3]{Tao Xue}
\author[3]{Jianmin Li}
\author[4]{Zongyi Wang}
\author[2]{Shaomin Chen}

\affiliation[3]{organization={Key Laboratory of Particle \& Radiation Imaging (Tsinghua University), Ministry of Education},
            city={Beijing},
            postcode={100084}, 
            country={P.R. China}}

\affiliation[4]{organization={School of Civil and Hydraulic Engineering, Huazhong University of Science and Technology},
            city={Wuhan},
            postcode={430074}, 
            state={Hubei},
            country={P.R. China}}

\cortext[1]{Corresponding author}

\fntext[1]{}


\begin{abstract}
	The Jinping Neutrino Experiment (JNE) will utilize approximately 3000 8-inch MCP-PMTs identified as GDB-6082 from North Night Vision Technology to detect neutrinos. To enhance the effective coverage of the JNE detector, mounting a custom-designed light concentrator on each PMT is a practical and economical approach. We measured angular responses of the concentration factor at four wavelengths in air medium for the concentrator with the selected cutoff angle of $70^\circ$. The measurements align with the Monte Carlo simulations. 
	Furthermore, our results indicate that these concentrators can improve the efficiency of light collection by 40\% under parallel illumination conditions. This enhancement results in a slight increase in transit-time spread, with the full width at half maximum (FWHM) increasing by less than 0.3 ns. We conclude that the developed light concentrators are highly suitable for the JNE. 
\end{abstract}



\maketitle

\section{Introduction}\label{sec:Introduction}
	The Jinping neutrino experiment (JNE)~\cite{Jinping:2016iiq} is the next-generation solar neutrino experiment at the China Jinping Underground Laboratory (CJPL) with a rock overburden of approximately 2,400 meters~\cite{Ma:2021uzi}. 
	As with other neutrino detectors~\cite{Super-Kamiokande:2002weg,SNO:1999crp,DayaBay:2012fng, Borexino:2008gab,IceCube:2016zyt, adrian2016letter}, the JNE will install photomultiplier tubes (PMTs) to detect photons emitted during particle interactions with matter and determine the particle's energy, position, and direction~\cite{Luo:2023reconstruction}. 
	Due to the extremely low probability of neutrinos interacting with matter, an ultralow background environment is essential to detect rare neutrino signals~\cite{wu2023performance}, including solar neutrinos~\cite{Xu:2022wcq, shao2023potential}, geoneutrinos~\cite{Bellini:2013wsa, wan2017geoneutrinos, wang2020hunting}, supernova neutrinos~\cite{DeGouvea:2020ang, wei2017discovery}, and neutrinoless double beta decay~\cite{Dolinski:2019nrj, fu2024comparison}. 
	The JNE is specifically being constructed to detect and study these neutrino signals.  
	
\begin{figure}[h]
		    \centering
		    \includegraphics[width=0.40\linewidth]{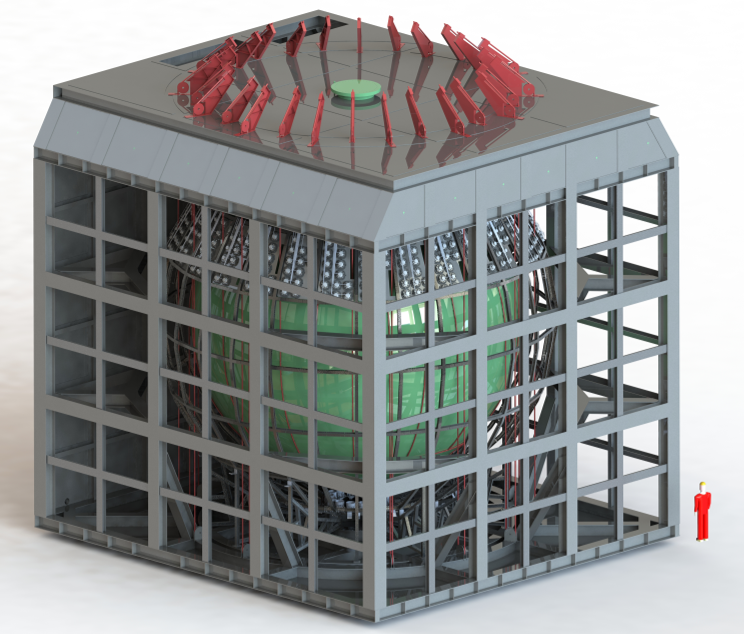}
		    \caption{The key structure of the JNE detector, which mainly consists of three parts: the inner acrylic sphere, the PMTs and associated supporting grid in the middle, and the outermost water shield.}
		    \label{fig:JNEDetector}
		\end{figure}

	The JNE detector (Fig.~\ref{fig:JNEDetector}) consists of three layers. The innermost layer is a transparent acrylic sphere, $4.98$ m in radius, containing the target medium. The middle layer is a spherical array of PMTs supported by a stainless steel frame $6.03$ m in radius, responsible for capturing the generated photons. The outermost layer is a water tank with dimensions of 14 $\times$ 12.9 $\times$ 13.2 m$^3$, which shields the inner volume from radioactivity in the external environment.
	The acrylic sphere is suspended within the detector using specially designed holding ropes to counteract gravitational or buoyancy forces~\cite{Wang:2024upf}. This sphere can be filled with various target materials in different experimental stages, including pure water, liquid scintillator~\cite{Guo:2017nnr}, and the aqueous solution of LiCl~\cite{Shao:2022yjc}. 

\begin{figure}[h]
		    \centering
		    \includegraphics[width=0.20\linewidth]{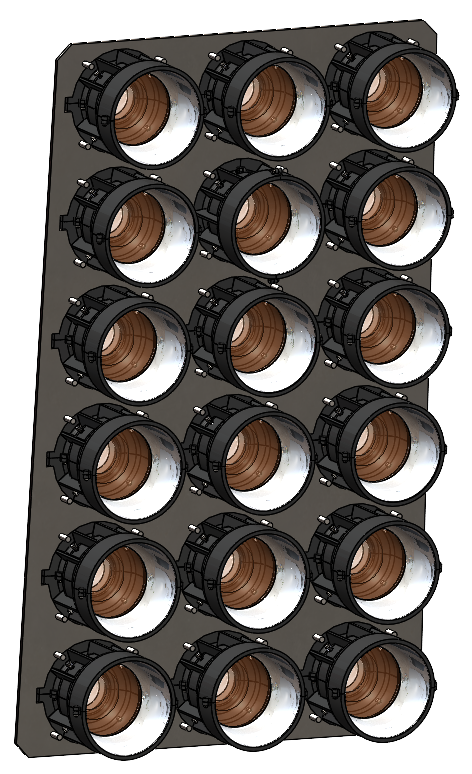}
		    \caption{The combination of PMT plus concentrator as the photon detecting device. } 
		    \label{fig:JNEconcentrator}
		\end{figure}

	The micro-channel plate (MCP) PMTs identified as GDB-6082~\cite{Zhang:2023ued} produced by North Night Vision Technology (NNVT) were selected as photon detection devices due to the high photon detection efficiency (DE) of up to 30\%. 
	Approximately 3000 MCP-PMTs will be deployed in the JNE detector. The energy resolution of the JNE detector is closely related to its overall photon collection efficiency, which is improved by equipping the PMTs with light concentrators, as shown in Fig.~\ref{fig:JNEconcentrator}. 
	The PMTs, together with the concentrators, are installed on a stainless steel plate according to a modular and zoned configuration. Depending on the specific shapes of each zone, each module contains a varying number of PMTs. Eventually, each module is positioned on the corresponding PMT grid. 

	The application of light concentrators on conventional PMTs in neutrino experiments and gamma-ray observatories is effective in achieving excellent physics results, notably in the cases of SNO, SNO+, Borexino~\cite{Lay:1996tp, SNO:2022qvw, Oberauer:2003ac}, HESS~\cite{bernlohr2003optical}, and VERITAS~\cite{radu2000design}. 
	Furthermore, OSIRIS~\cite{Loo:2023kij} and KamLAND-Zen~\cite{shirai2017results} have also proposed incorporating concentrators into their detectors. In the realm of $\gamma$-ray astronomy, mounting light concentrators on SiPMs represents an additional viable strategy~\cite{Aguilar:2014oba, Krizan:2024ydm}.
	
	This paper describes the development and characterization of the concentrator. 
	We have formulated a concentrator with a wide field of view and a high geometric collection efficiency based on mathematical and simulated studies~\cite{Zhi:2017tbk}, which improves the traditional string method and effectively increases the photon collection efficiency. 
	The concentrator functions by expanding the entrance aperture of the PMT and reflecting the incoming light onto the PMT's photocathode. 
	As PMTs are expected to cover about 25\% of the solid angle, a circular opening light concentrator is chosen to increase the effective coverage rather than a hexagonal one. 
	On the other hand, the concentrator can shield light beyond the region of the cut-off angle, e.g., radioactivities from neighboring PMTs.   
	
	The structure of the article is as follows. Section \ref{sssec:design} introduces the theoretical design scheme of the light concentrator; Section \ref{sssec:material} discusses the selection of manufacturing materials and their mechanical performance; Section \ref{sssec:exptesting} elaborates on the validation of the concentration factor via the experiment and Monte Carlo (MC) simulations; Section~\ref{sssec:discussion} and \ref{sssec:conclusion} contain the discussion and conclusion.

\section{The design of the concentrator}\label{sssec:design}
	The design of the reflector surface of the concentrator takes into account the maximization of the light reflection efficiency and good contact with the photocathode, and the characteristics of the photocathode, such as DE and transit time spread (TTS), to ensure that the combination of PMT plus concentrator is in the optimal working condition. 
	Although adding concentrators around the PMTs increases the diameter of the entrance aperture, it reduces the observed field of view.
	Therefore, it is imperative to determine the effective detection volume of the experiment to ensure that it can fully cover the acrylic sphere. In addition, the economy and practicality of the concentrator are also essential features. When choosing a simple and cost-effective design, it is necessary to ensure the safety and stability of the overall structure and adopt a convenient installation method to shorten the construction period. 

	\subsection{The MCP-PMT of GDB-6082}
		The MCP-PMT (GDB-6082) comprises a glass ellipsoid, a glass neck, and a plastic base. The three semi-axes of the ellipsoid are 10.15, 10.15, and 7.36 cm in length, with the photocathode distributed on the front part above the ellipsoid's equator. The ellipsoid has a highly reflective aluminum-coated layer below the equator and around the neck, which can effectively reflect incident light back to the photocathode. The base is made of ABS plastic and houses a voltage divider module and is filled with waterproof and oil-resistant materials. Generally, when the voltage is around 1700 V, the photoelectron (PE) gain of the PMT can reach an order of magnitude of 10$^7$.
		
\begin{figure}[h]
			    \centering
			    \includegraphics[width=0.48\linewidth]{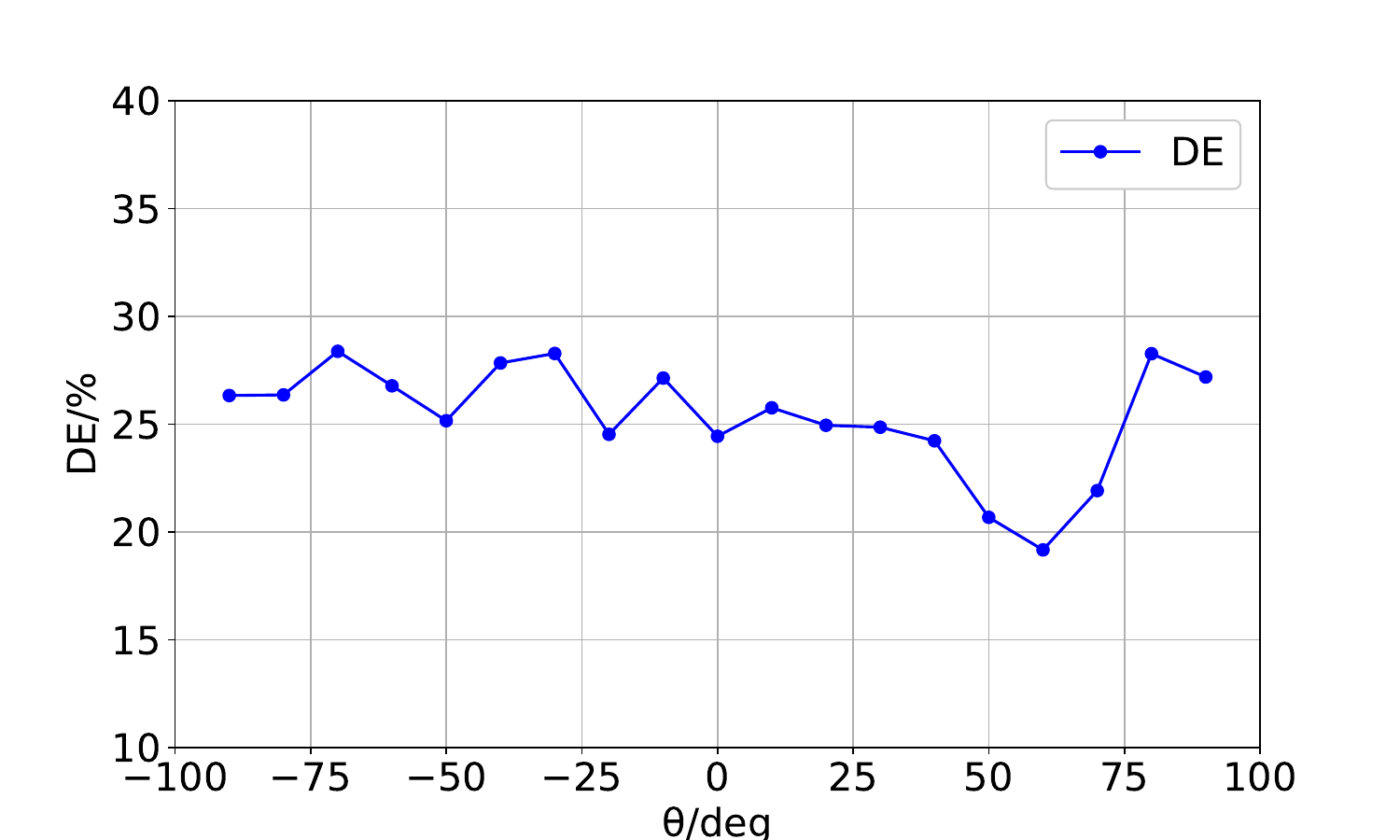}
			    \includegraphics[width=0.48\linewidth]{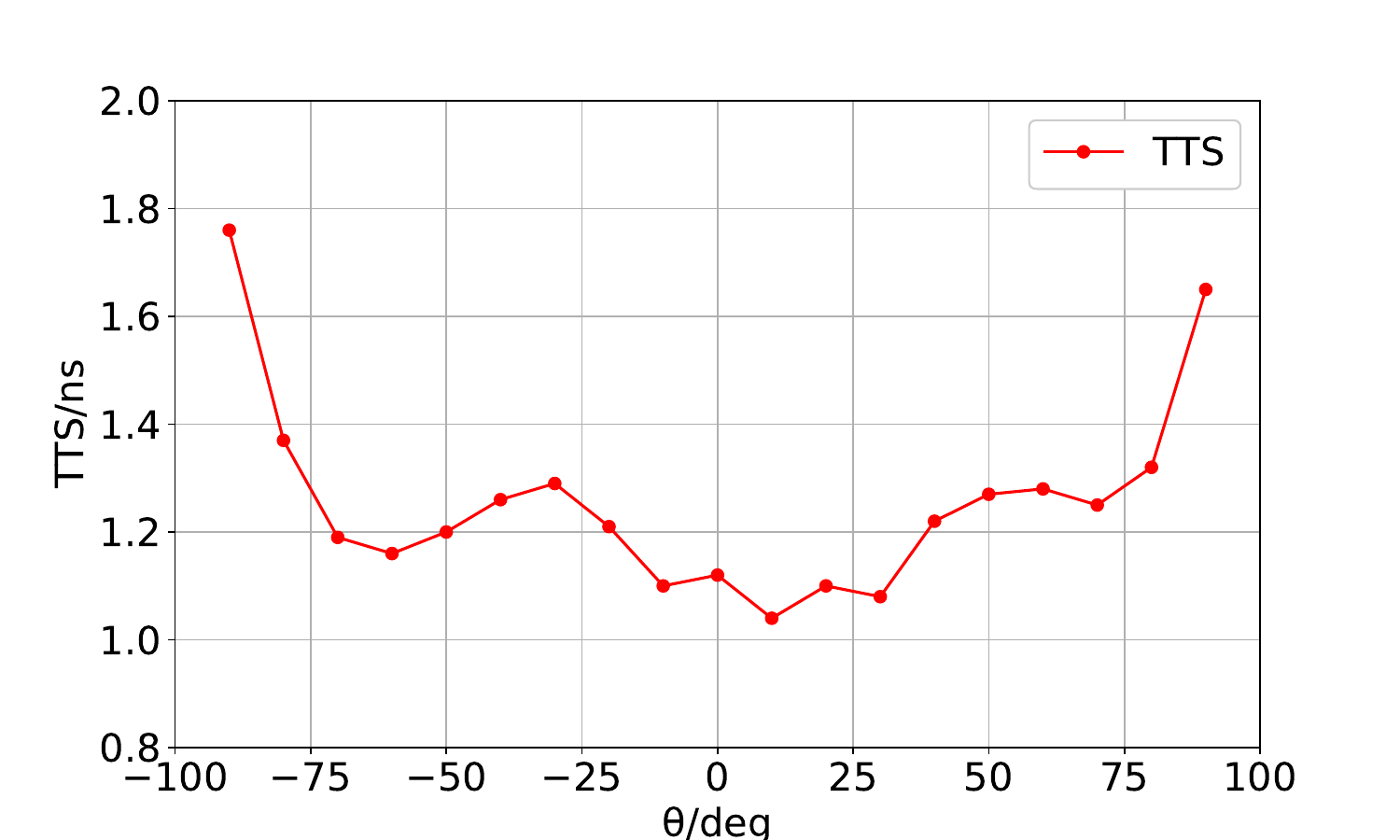}
			    \caption{The variation of DE (left) and TTS (right) of MCP-PMT (GDB-6082) with respect to the zenith angle $\theta$. Here, $\theta$=0$^{\circ}$ corresponds to the front vertex of the PMT ellipsoid, while $\theta$=$\pm 90^{\circ}$ corresponds to the equator of the ellipsoid. A significant increase in the vicinity of the equator is observed for the TTS scanning. } 
			    \label{fig:PMTTTS}
			\end{figure}

		The performance of the PMT has been extensively studied in various aspects, including DE, TTS, dark count rates (DCR), the probability of pre- pulses and after-pulses, the single-electron response (SER) and a long tail in the charge distribution~\cite{Zhang:2023ued}, which is parameterized with the Gamma-Tweedie mixture model and reported in ~\cite{Weng:2024tjs}. In addition, the TTS is unevenly distributed with respect to the zenith angle ($\theta$) in the ellipsoid (see Fig.~\ref{fig:PMTTTS}) and increases significantly near the equatorial plane of the ellipsoid, while the correlation between the DE and the elevation angle is relatively small. This indicates that the transit time of the photoelectrons emitted near the equator is longer, which leads to an increased TTS and affects the accurate measurement of a particle's energy-deposition position within the detector. We could mask this equatorial region using the light concentrator. 

	\subsection{Light bowl}
		The light bowl used is designed based on a 3D-modified string method \cite{Zhi:2017tbk}. In this method, the key parameters that determine the shape of the light bowl are the light exit aperture and the cutoff angle. Higher collection efficiency of photons requires a larger entry aperture of a light bowl, and thus a larger exit aperture. So, it is best to set the exit aperture of the light bowl to coincide with the equatorial plane of the PMT photocathode. However, as shown in Fig.~\ref{fig:PMTTTS}, for the MCP-PMT used in JNE, the TTS around the equatorial plane of the photocathode is quite high, leading to a poorer position resolution and discrimination between scintillation and Cherenkov light, which is essential for particle identification. As a result, only the part with a zenith angle less than $|\pm 80^{\circ}|$ of the photocathode is used, where a TTS less than 1.4 ns is guaranteed. The exit aperture of the light bowl is thus a circle with a radius of 9.98 cm and 1.76 cm higher than the equatorial plane of the photocathode. And the PMT also benefits from being tightly clamped by the concentrator. Further details are provided in Section.~\ref{sssc:assembly}.
		
\begin{figure}[h]
			        \centering
			        \includegraphics[width=0.7\linewidth]{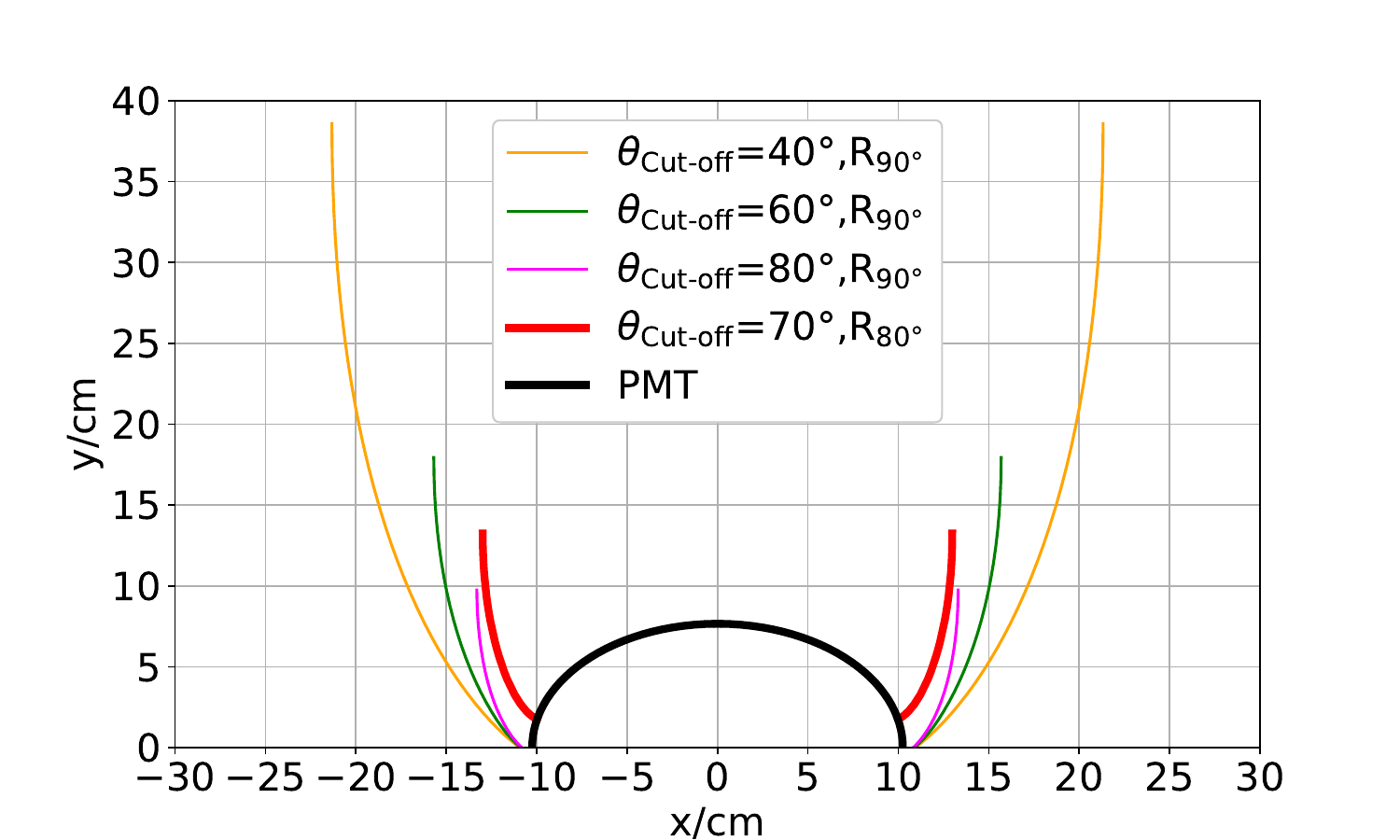}
			        \caption{The cross-section of the PMT and concentrator with different cut-off angles $\theta_{\text{Cut-off}}$ in the xy plane. 
			        Here, R$_{90^{\circ}}$ represents that its exit aperture is aligned with the equator of the PMT, while R$_{80^{\circ}}$ indicates that its exit aperture is located at $\pm 80^{\circ}$.
			        }
			        \label{fig:cut-off-angle}
			\end{figure}

		The modified string method ensures that most rays with an incident angle less than the cut-off angle reach the PMT photocathode either directly or after a few reflections. As illustrated in Fig.~\ref{fig:cut-off-angle}, a higher cut-off angle results in a lower and narrower light bowl, allowing more rays from a broader range to be collected. This leads to a larger effective detection volume, which can be calculated as detailed in Table.~\ref{tbl:cut-off}. 
		
\begin{figure}[h]
			        \centering
			        \includegraphics[width=0.55\linewidth]{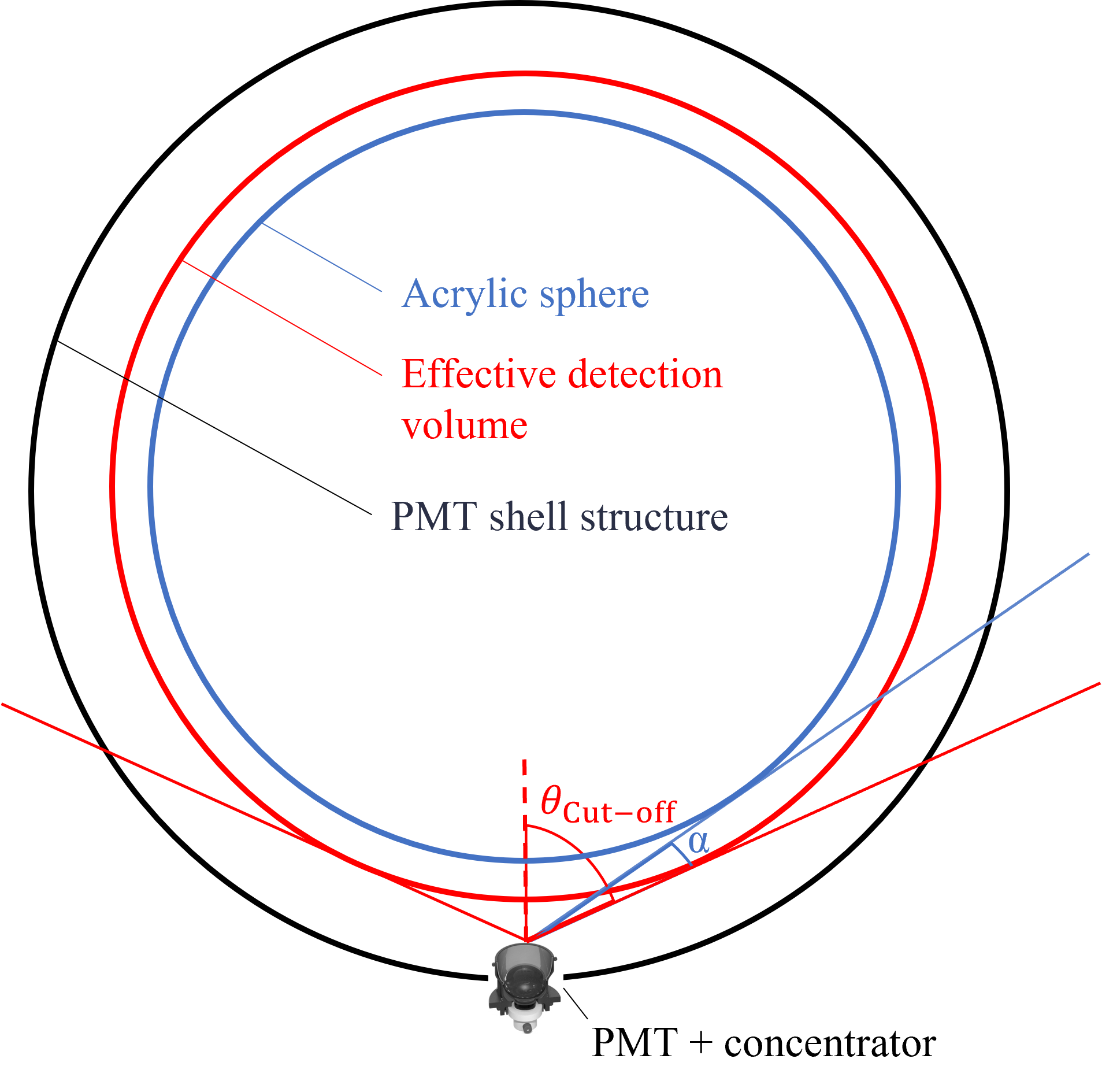}
			        \caption{The cut-off angle $\theta_{\text{Cut-off}}$, installation error angle $\alpha$ and effective detector volume of a single PMT equipped with a concentrator. Within the error range, the acrylic spherical detector of JNE is completely contained within the effective detection volume.}
			        \label{fig:effective-detection-volume}
			\end{figure}

		In this context, light bowls of the concentrator with cut-off angles in the range $60^{\circ}$ $-$ $70^{\circ}$ are compared based on several parameters: the radius of the effective detection volume (R$_{\text{eff}}$), the maximum allowed installation error ($\alpha$), the ratio of the entrance aperture to the exit aperture area of the PMT (R$_{\text{S}}$), the average photon transition time (T$_{\text{tt}}$) defined as the mean time difference from passing through the entrance aperture to the photocathode, and T$_{\text{tts}}$, measured using the full width at half maximum. Note that the TTS of the light concentrator is less than $0.3$ ns, which has a negligible effect on the combined TTS of the PMT plus concentrator since $\sqrt{(1.4)^2 + (0.3)^2} \approx 1.4$ ns assuming that the two TTSs are independent.  The $\alpha$ in Fig.~\ref{fig:effective-detection-volume} shows the maximum angular tolerance in which the effective detection volume just covers the liquid scintillator.
		Table~\ref{tbl:cut-off} summarizes the results. Given that the angle error during PMT installation may reach a maximum of 10$^{\circ}$, we have chosen the light bowl with the cut-off angle of 70$^{\circ}$, as shown in Fig.~\ref{fig:cut-off-angle}.

			\begin{table}
				\caption{The key parameters of the different light bowls obtained through calculation and simulation include: the cut-off angle $\theta_{\text{Cut-off}}$, the radius of the effective detection volume R$_{\text{eff}}$, the maximum allowable installation error angle $\alpha$, the ratio of the entrance aperture to the exit aperture area of the PMT R$_{\text{S}}$, and the transmission time T$_{tt}$ and the total transmission time spread T$_{tts}$ brought about by the use of the concentrator. During the calculation and simulation process, the exit aperture is fixed at $\theta$ = $\pm 80^\circ$. When conducting the simulation analysis for T$_{tt}$ and T$_{tts}$, the incident angle of the light is a random value ranging from 0 to $\theta_{\text{Cut-off}}$.}
				\label{tbl:cut-off}
				\begin{tabular*}{\tblwidth}{@{}CCCCCCC@{}}
					\toprule
					$\theta_{\text{Cut-off}}$ [${}^{\circ}$] & R$_{\text{eff}}$ [m] & $\alpha$ [${}^{\circ}$] & R$_{\textbf{S}}$ &  Average T$_{tt}$ [ns] & T$_{tts}$ [ns] \\
					\midrule
					            70.0 & 5.43 & 10.22 & 1.60  & 0.47 & 0.27 \\
					            69.0 & 5.39 & 9.15 & 1.63   & 0.49 & 0.23 \\
					            68.0 & 5.35 & 8.08 & 1.66   & 0.50 & 0.27 \\
					            67.0 & 5.32 & 7.00 & 1.68  & 0.52 & 0.23 \\
					            66.0 & 5.27 & 5.92 & 1.71  & 0.53 & 0.23 \\
					            65.0 & 5.23 & 4.84 & 1.74  & 0.55 & 0.27 \\
					            64.0 & 5.19 & 3.76 & 1.77  & 0.57 & 0.23 \\
					            63.0 & 5.14 & 2.67 & 1.81  & 0.59 & 0.23 \\
					            62.0 & 5.09 & 1.57 & 1.84 & 0.60 & 0.23 \\
					            61.0 & 5.04 & 0.48 & 1.88  & 0.63 & 0.20 \\
					            60.0 & 4.99 & 0.31 & 1.92  & 0.65 & 0.23 \\
					\bottomrule
				\end{tabular*}
			\end{table}

	\subsection{Assembly structure}\label{sssc:assembly}
\begin{figure}[h]
			    \centering
			    \includegraphics[width=0.65\linewidth]{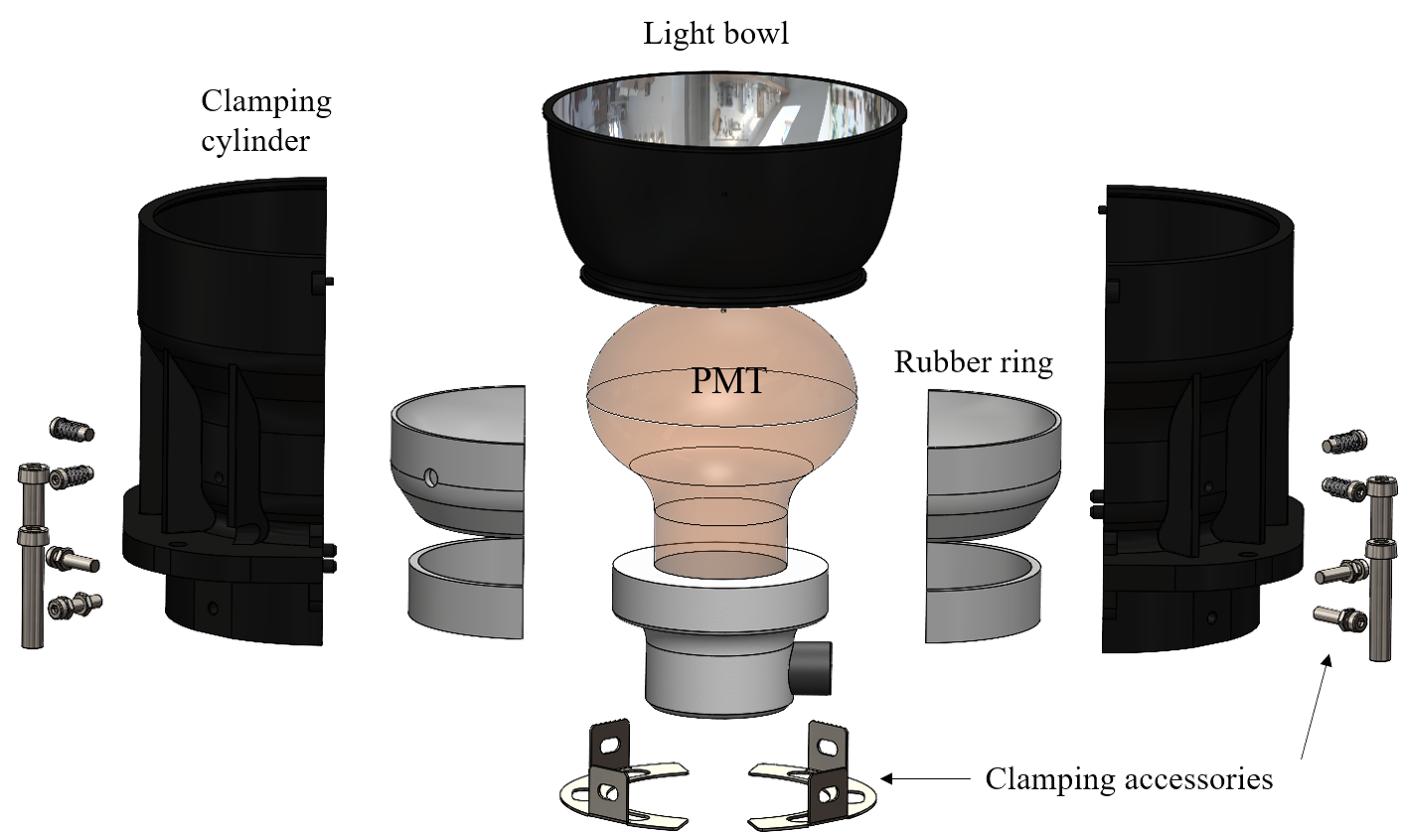}
			    \includegraphics[width=0.33\linewidth]{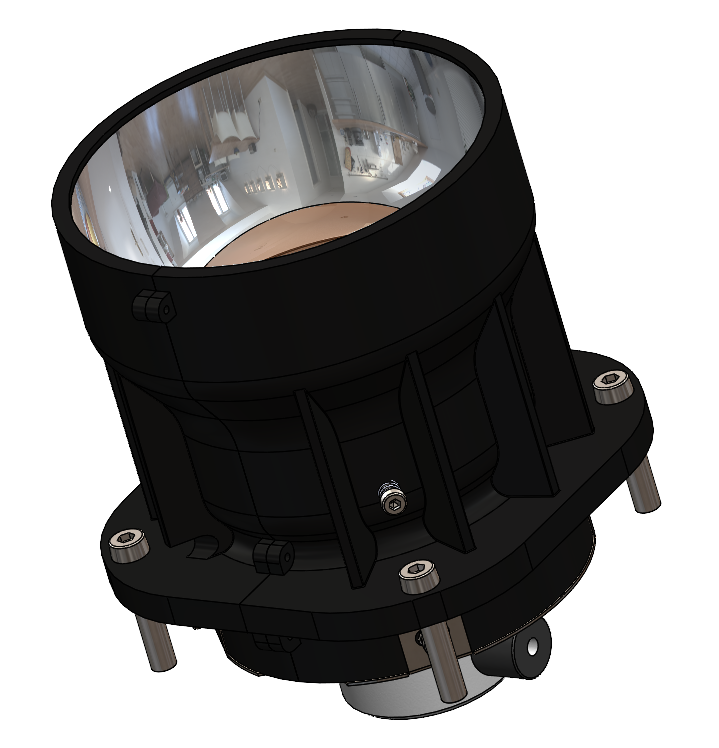}
			    \caption{The disassembly and assembly diagrams of MCP-PMT and concentrator. The structure of the concentrator mainly consists of clamping cylinder, a light bowl, silicone rubber rings and clamping accessories. } 
			    \label{fig:StructureOfConcentrator}
			\end{figure}

		A specialized assembly system has been designed to properly secure the light bowl at the top of the PMT and to facilitate the installation of both the PMT and the light bowl on the grid shell of the JNE detector. 
		The basic structure of the assembly scheme presented in this paper consists primarily of a clamping cylinder, a light bowl, silicone rubber rings, and clamping accessories, as shown in Fig.~\ref{fig:StructureOfConcentrator}.
		
		The clamping cylinder features a symmetric design, ensuring a uniform force distribution on both sides. This design facilitates processing and production with the same set of molds. Furthermore, this structure can be directly connected to the detector cover plate using four M16 screws, allowing for a secure and quick installation.  
		
		The light bowl is a key working component of the light concentrator. Not only does it provide a reflective surface, but the curved structure at its bottom also ensures a close fit with the PMT, achieving stable fixation. During the design process, the shape tolerance of the PMT sphere was fully considered, achieving a seamless connection. 
		
		The silicone rubber ring, as the main clamping component of the light concentrator, is made of moderately hard silicone rubber material and is specially designed for the non-working area below the equator of the PMT sphere. The ring works in conjunction with the clamping cylinder and the light bowl to ensure secure clamping of the PMT head. 
		
		The clamping accessories include the fastening screws and nuts that assist in assembling and securing the overall structure. 

\section{Material }\label{sssec:material}
	\subsection{Selection of material}
		In selecting the material for the light concentrator, we referred to research on the SNO experiment~\cite{SNO:1999crp} and the materials used in the PMT packaging structure~\cite{Zhang:2023ued}, and ultimately chose high-performance ABS plastic.  
		
		ABS plastic offers several advantages: Its high strength allows it to support the weight of the PMT and its associated concentrator while withstanding the buoyancy the PMT experiences in water. Second, its excellent wear resistance and surface stress ensure that the shape of the light concentrator remains consistent over time. In addition, ABS plastic has stable chemical properties and is resistant to water and mineral oil, helping to prolong its service life. Finally, as a plastic material, its radioactive background can be well under control, thereby reducing interference to the signal of interest. 
		
		The mechanical properties of ABS plastic were measured using a uniaxial tensile test based on the guideline of  ASTM D638-14 ~\cite{astm2014standard}. The test recorded the stress-strain curve, which revealed the elastic modulus, tensile strength, and fracture behavior of the material. The results show that ABS has an average tensile strength of $37.14$ MPa and an elastic modulus of $2392.92$ MPa.
		The combined structure of PMT and the associated concentrator was modeled and sectioned into meshes using HyperMesh~\cite{altair2017hyperworks} , and finite element analysis was performed with ABAQUS~\cite{Hibbitt:1984abaqus} to analyze the mechanics of the combined structure at four positions at different depths under water. The results are summarized in Table~\ref{table:ABS_Plastic}. The calculations show that the maximum principal stress is $0.60$ MPa, which is about $1.6\%$ of the tensile strength of ABS, demonstrating that ABS plastic is secure even at a 12-meter depth of water and is expected to remain stable for long-term operation.  

			\begin{table}
			\caption{Summary of mechanical results of finite element analysis at four positions of different depths under water. The second column lists the depth of the position, and the third column shows the maximum principle stress.}
			\label{table:ABS_Plastic}
			\begin{tabular*}{\tblwidth}{@{}CCC@{}}
			\toprule
			Position & Depth (m) & Maximum principle stress (MPa) \\
			\midrule
			                A & $1.77$ & $0.56$\\
			                B & $6.03$ & $0.40$\\
			                C & $10.29$ & $0.49$ \\
			                D & $12.03$ & $0.60$ \\
			\bottomrule
			\end{tabular*}
			\end{table}

	\subsection{Reflecting surface}
		Since aluminum has a strong reflectivity for light of different wavelengths, we selected it as the reflective layer of the light bowl. The reflectivity of aluminum in the visible light band (400-700 nm) can reach 80\%, and it also maintains a high reflectivity in the ultraviolet region (part of the light-sensitive area of PMT\cite{Zhang:2023ued}).  
		
\begin{figure}[h]
	            \centering
	            \includegraphics[scale = 0.5]{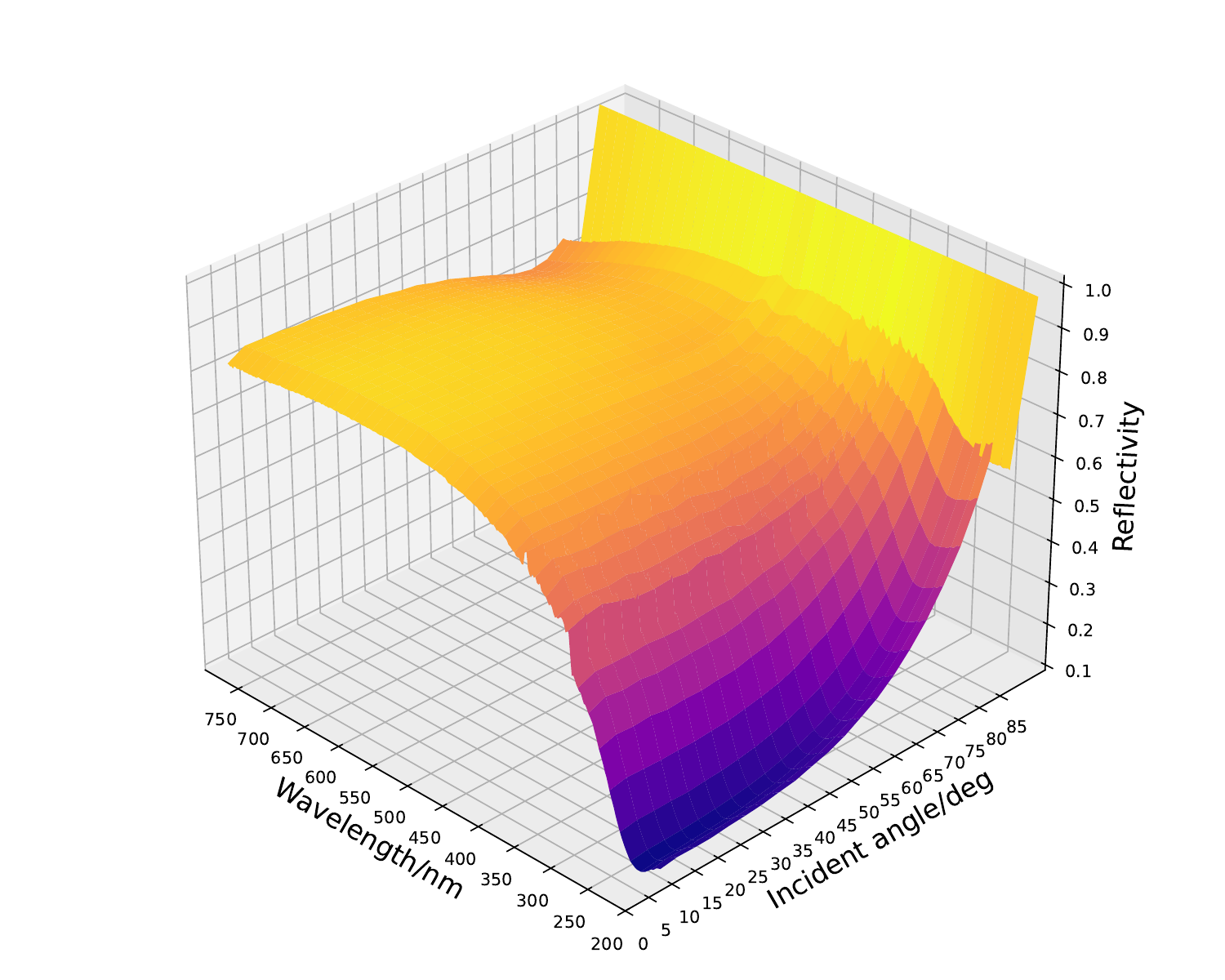}
	            \caption{The reflectivity of the coated aluminum versus wavelength and incident angle. } 
	            \label{fig:concentratorref}
	        \end{figure}

		To achieve efficient reflective performance, we use vacuum coating technology to evenly coat aluminum on the light bowl made of ABS plastic and ensure that its thickness is greater than 50 nm. The surface roughness and oxidation can affect the reflectivity of the aluminum film. The reflectivity of coated aluminum versus wavelength and incident angle, as shown in Fig.~\ref{fig:concentratorref} is measured using a UV/Vis/NIR spectrophotometer.  Note that the reflectivity at $90^\circ$ is not measured, but is assumed to be one as per the Fresnel law.  
        
\section{Validation of the concentration factor}\label{sssec:exptesting}
	Through the collaborative efforts of the R\&D team, we successfully fabricated the concentrator and completed its assembly with the PMT. To assess the performance of the concentrator, we designed a dedicated test facility and conducted detailed experimental tests. We compared the experimental results with those obtained from Geant4 simulations. The consistency between the measured and simulated results demonstrated that the concentrator significantly enhanced the photon collection efficiency.

	\subsection{The test facility}
\begin{figure}[h]
			    \centering
			    \includegraphics[width=0.75\linewidth]{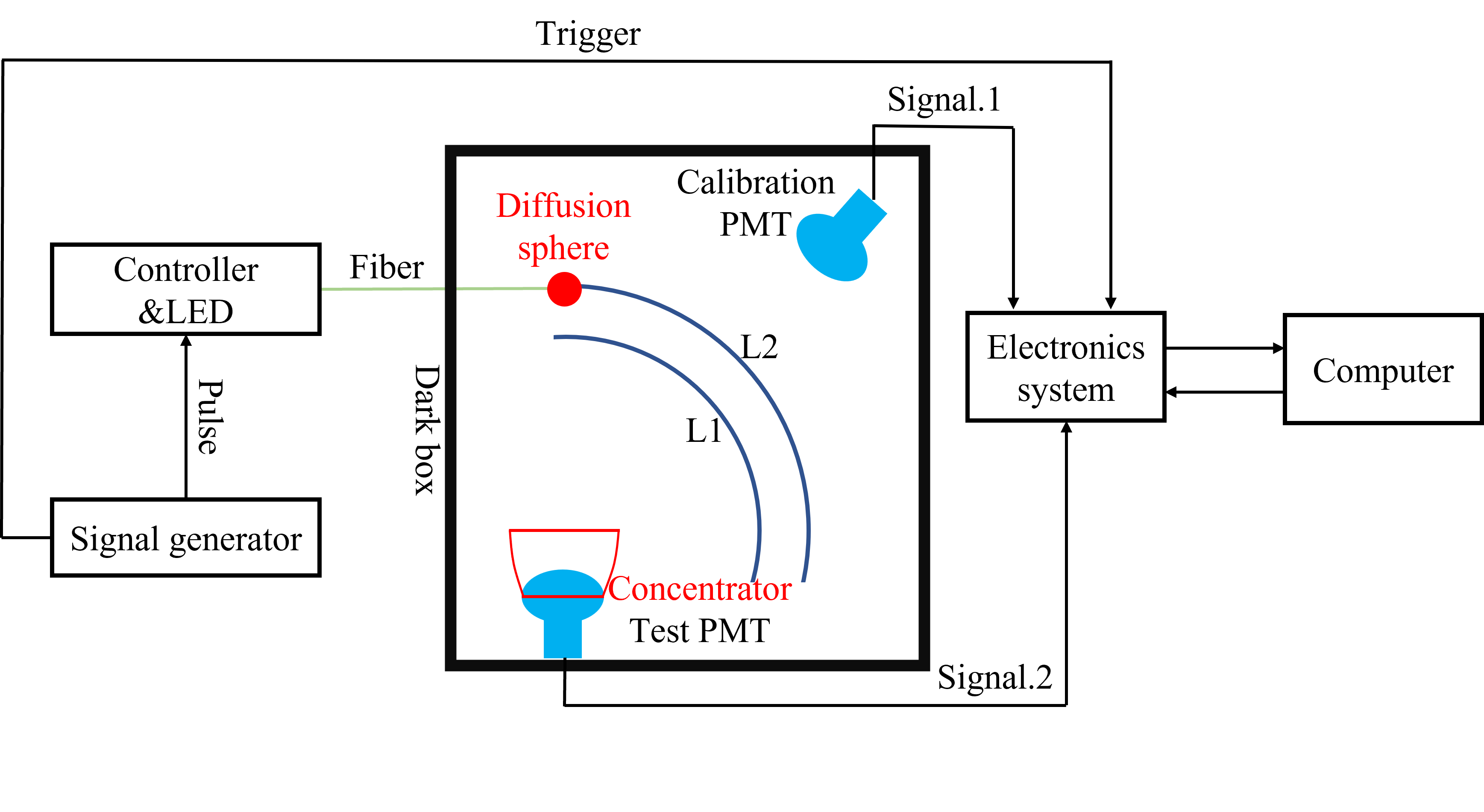}
			    \caption{The test facility, including a dark box, a light source system, a readout electronics system and a computer.
			    }
			    \label{fig:DarkBox}
			\end{figure}

		As shown in Fig.~\ref{fig:DarkBox}, the test facility comprises a dark box, a light source system, a readout electronics system, a computer, and two identical 8-inch MCP-PMTs (GDB-6082), one for testing and one for calibration. Installation and removal operations can be conveniently carried out on the test PMT, which is then used to measure the light concentrator's performance.

		The black box is used to shield external light to ensure the accuracy of the experimental measurements. It measures 1.53 $\times$ 1.43 $\times$ 0.37 m$^3$ and is made of 1 cm thick black opaque acrylic plates.
		We frosted the inner sides of the acrylic plates to reduce the impact of internal light reflection on the test results.
		Two quarter-circular rails are installed at the bottom of the dark box, namely L1 and L2, allowing the light source to slide along the rails. These two rails share a common center, which is the center of the entrance aperture of the light bowl. The radii of L1 and L2 are 50 and 75 cm, respectively. The stainless steel support columns on the rails ensure that the light source remains at the same horizontal level as the center of the light bowl. 
		
		The light source system consists of LEDs, a signal generator, a controller, a diffusion sphere, and an optical fiber. Its working principle is as follows: The signal generator generates pulse waveforms and outputs them to the controller. The controller then drives the LEDs to emit light, transmitting it through the optical fibers to the diffusion sphere inside the dark box for uniform emission. 
		The LEDs in this system emit light at multiple monochromatic wavelengths, including 365, 415, 465, and 480 nm. The half-width of each wavelength is approximately 10 to 20 nm. 
		We have configured the signal generator with different duty cycles to optimize the performance of the LEDs at varying wavelengths.
		Specifically, the duty cycle for the 365 nm LED is set at about 3.5\%, the duty cycles of the 465 nm and 480 nm LEDs are 2.0\%, and the duty cycle of the 415 nm LED is only 0.2\%. The pulse frequency generated by the signal generator is 100 kHz. The selected optical fibers are commercial-grade single-mode optical fibers, which can provide stable optical transmission performance. The diffusion sphere is made of polytetrafluoroethylene material and has a diameter of 5 cm. 
		
		The readout electronics system for this device was independently designed and manufactured by the R\&D team. The waveform digitization board \cite{Yang:2024qqe} comprises a system motherboard and an FPGA mezzanine card (FMC). 
		The FMC utilizes AD9695 from Analog Devices, which is a 1000 MSPS, 14-bit JESD204B ADC. 
		It features two JESD204B channel outputs with a channel rate of up to 10 Gbps. Using DC input coupling, the input single-ended signal is converted to a differential signal through the fully differential amplifier LMH6654 from TI. The AD5686 from Analog Devices is used to bias the input signal to fully utilize the full-scale dynamic range of the AD9695, with a nominal value of 1.6 V. The motherboard of the system is a 6U CPCI standard module. It combines two FMC high pin count (HPC) connectors with an FPGA. Each FMC connector is connected to the FPGA through 58 LVDS signals and 8 pairs of high-speed serial transceiver signals. The theoretical bandwidth of each channel is 16.3 Gbps. The trigger input/output and the external reference clock input are designed on the front panel of the waveform digitizer board. A 64-bit DDR4 SDRAM is also designed to buffer real-time data. Additionally, a self-defined software module, running on ZYNQBee2 board~\cite{xue2018design} equipped with Xilinx's ZYNQ7010 core, is used for slow control and online firmware updates.
		
		A computer is required to control the electronic system, regulate the high voltage of the PMT, and collect and store data, in addition to the above equipment.   

	\subsection{Data analysis}
		\subsubsection{Waveform analysis} 
\begin{figure}[h]
			            \centering
			            \includegraphics[scale = 0.30]{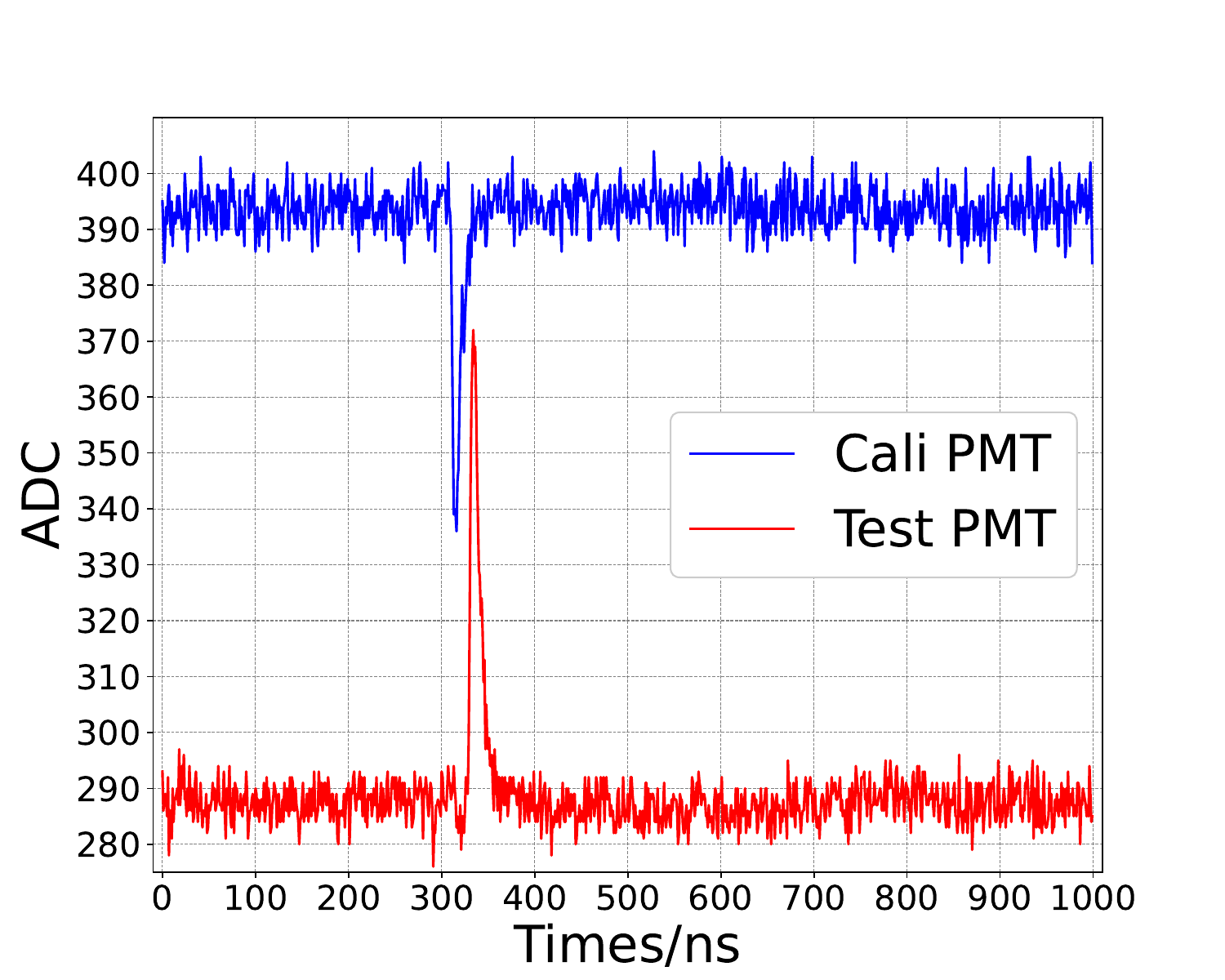}
			            \includegraphics[scale = 0.30]{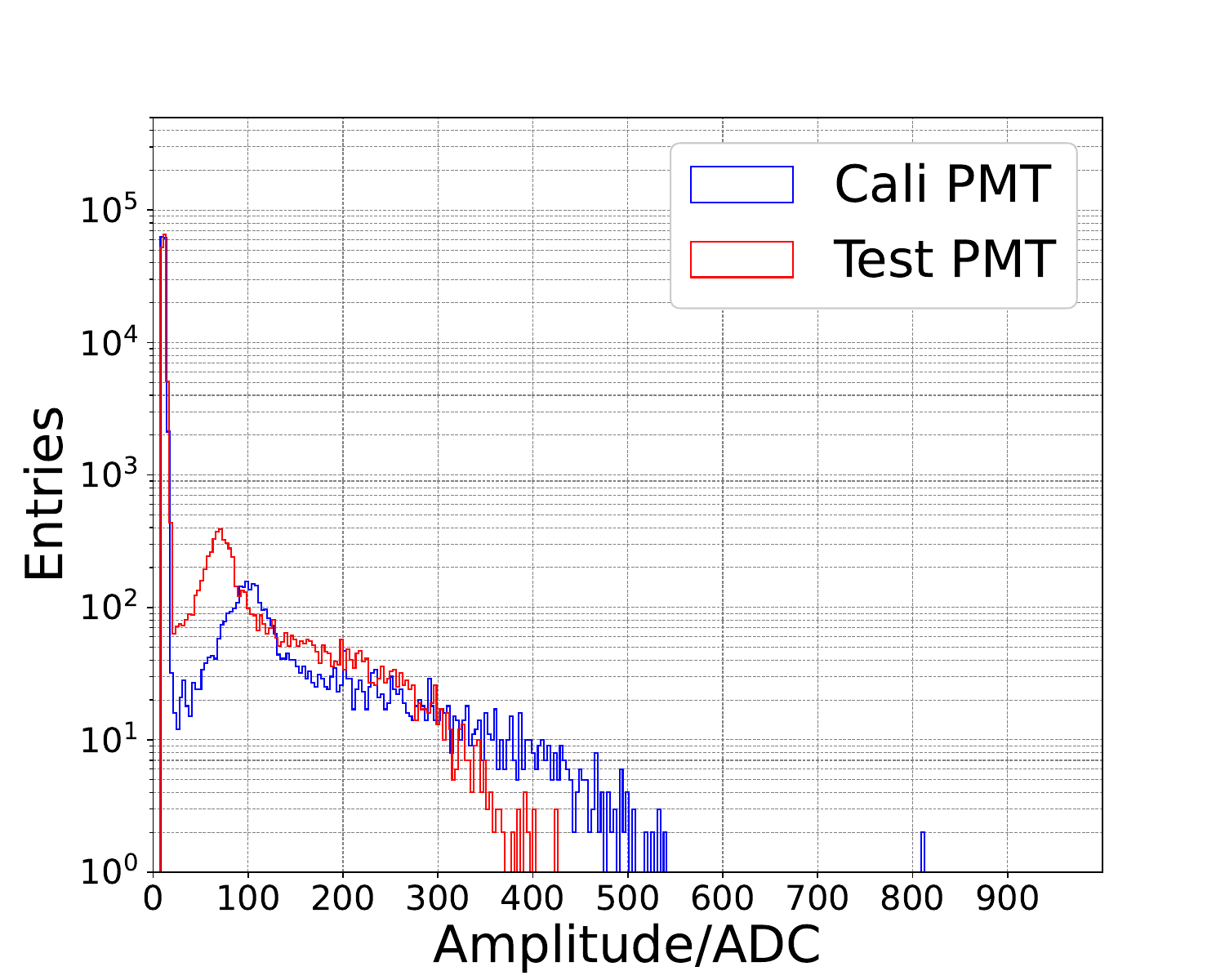}
			            \caption{Left: two typical waveforms from the calibration PMT and the test PMT. Right: amplitude distribution of photoelectron peaks from the calibration PMT and the test PMT. The blue (red) histogram corresponds to the calibration (test) PMT. Data shown as an example come from a certain run. The difference in the amplitude distributions between the test and calibration PMT is due to their gain difference, which does not affect the conclusions of this study.}  
			            \label{fig:OriginalWave}
			        \end{figure}

\begin{figure}[h]
			            \centering
			            \includegraphics[scale = 0.30]{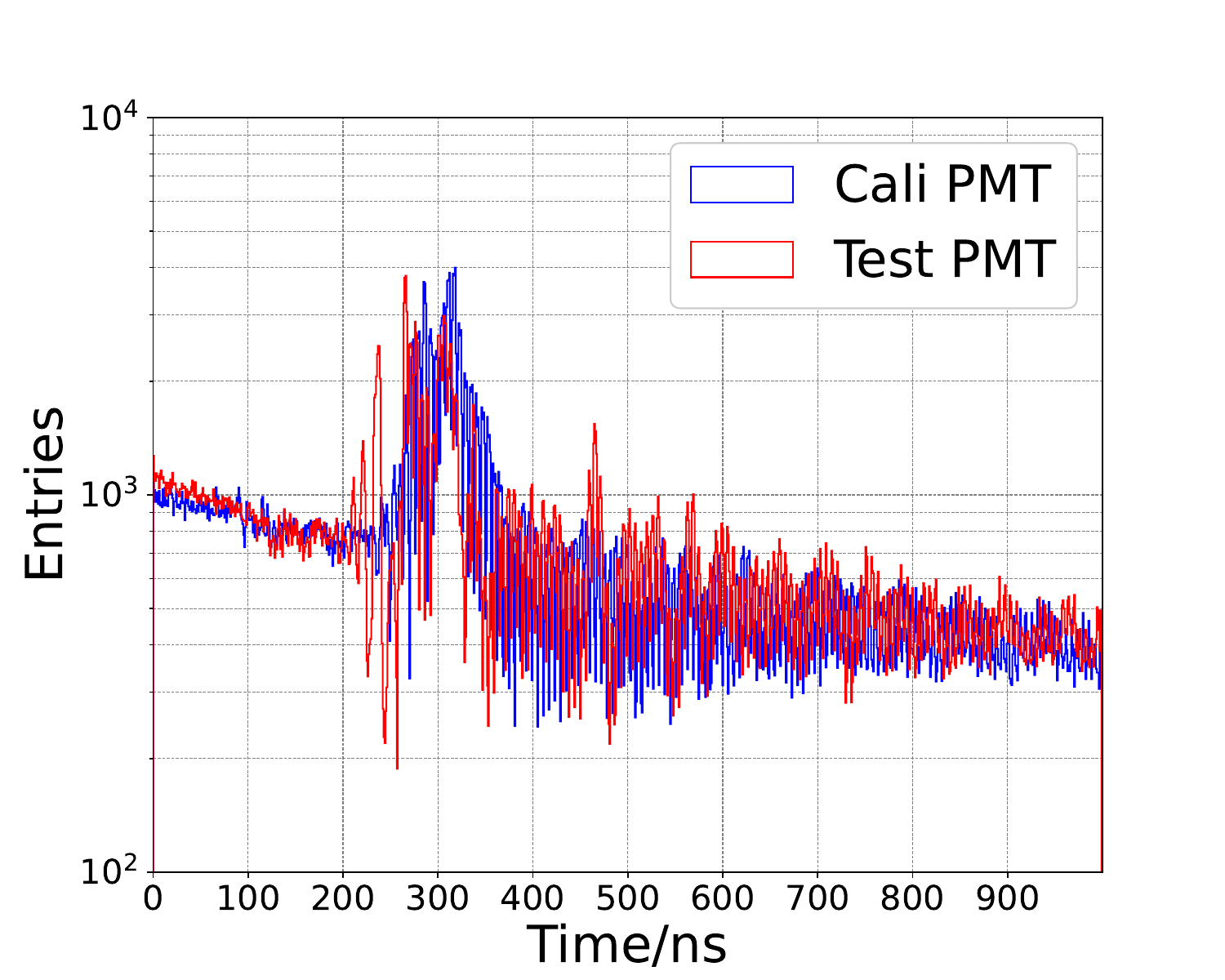}
			            \includegraphics[scale = 0.30]{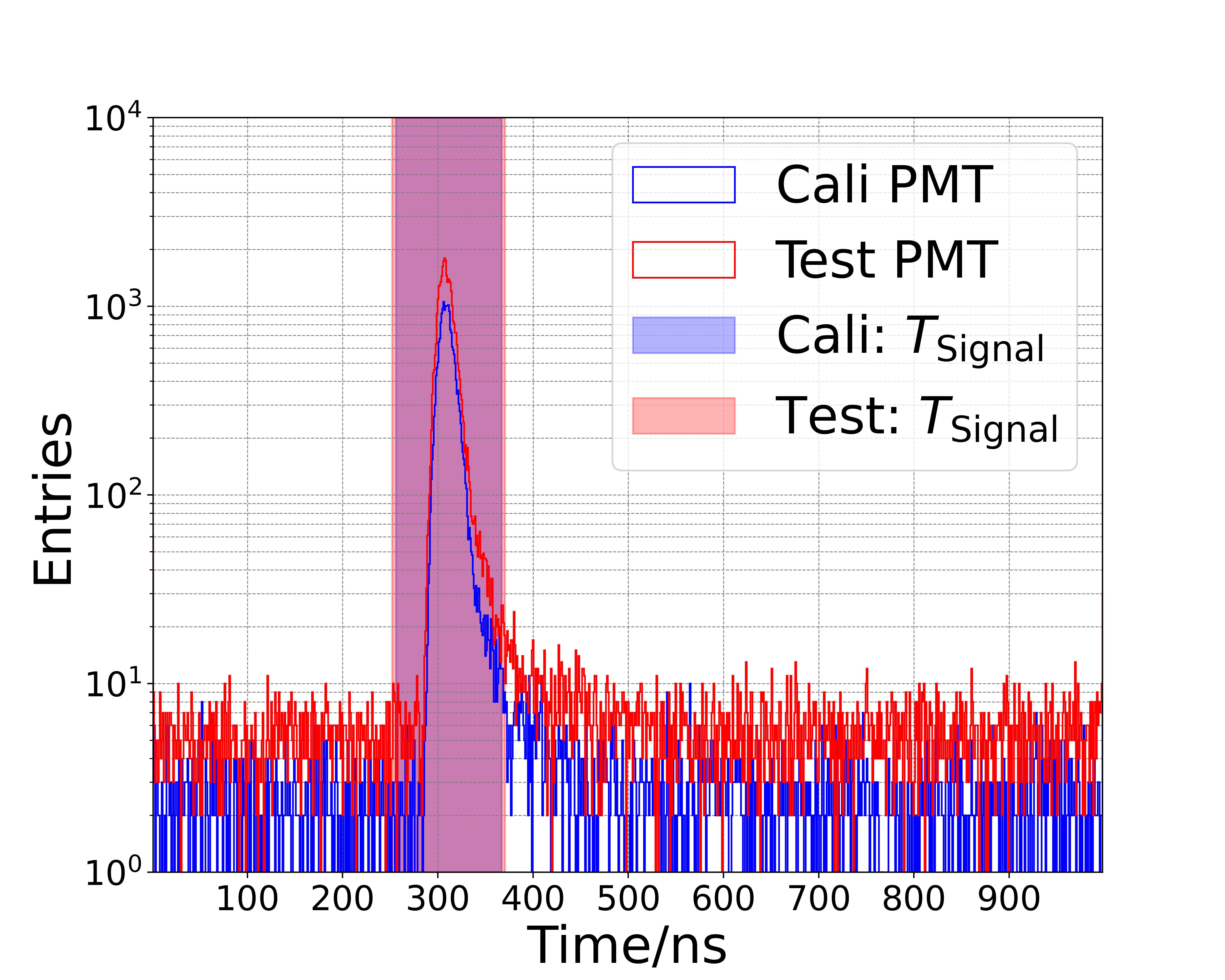}
			            \caption{Time distribution of photoelectrons peaks from the calibration PMT and the test PMT before (left) and after (right)  the pedestal cut. Blue (red) histogram represents the calibration (test) PMT. The shadowed regions represent the signal windows ($T_\mathrm{Signal}$). Data shown as an example come from a certain run.}  
			            \label{fig:WaveformA}
			        \end{figure}

			The left panel of Fig.~\ref{fig:OriginalWave} shows two typical waveforms of the calibration PMT and the test PMT. The waveforms were captured within a 1 us time window with a 1 GSPS sampling frequency. The baseline is subtracted to obtain the amplitude and time distributions of the photoelectron peaks.  A noise pedestal peak is observed in the right panel of Fig.\ref{fig:OriginalWave}, which can introduce significant noises, as shown in the left panel of Fig.\ref{fig:WaveformA}. 
			        
			To eliminate the effect, we only select peaks whose amplitudes are greater than $\mu_\mathrm{ped} + 5 \sigma_\mathrm{ped}$, where $\mu_\mathrm{ped}$ and $\sigma_\mathrm{ped}$ are fit from those amplitudes less than $30$ ADC using a Gaussian. The threshold of $\mu_\mathrm{ped} + 5 \sigma_\mathrm{ped}$ leads to less than $1.0\%$ signal loss, which is negligible, since the peak-to-valley ratio of the amplitude distribution can be as large as 5 $-$ 7. After applying this selection, we have obtained the time distribution of the photoelectron peaks, as shown in the right panel of Fig.~\ref{fig:WaveformA}. A Gaussian function $G(t_\mathrm{signal}, \sigma_\mathrm{signal})$ is used to fit and determine the signal window $[t_\mathrm{signal} - 3\sigma_\mathrm{signal}, t_\mathrm{signal} + 3 \sigma_\mathrm{signal}]$, from which the number of non-zero hits is counted. The contribution of dark noise hits is then subtracted based on estimations derived from the flat sideband beyond the signal window. 

		\subsubsection{Concentration factor}
		    The number of non-zero hits $N^{\mathrm{hit}}$ is the total number of hits with peak times falling within the signal window with $N^{\mathrm{total}}$ observations. The DCR is estimated as the rate of the hits with the peak time falling within the time window $[40$ ns, $t_\mathrm{signal} - 3\sigma_\mathrm{signal}]$, where only noise is expected and no signals should appear. The contribution from dark noise:
            \begin{equation}
                N^\mathrm{noise}
                =6\sigma_\mathrm{signal}N^{\mathrm{total}}\times\mathrm{DCR}
            \end{equation}
            in the width of signal window $6\sigma_{\mathrm{signal}}$ is subtracted in the trigger ratio $R$ as
			    \begin{equation}
			        R=(N^{\mathrm{hit}}-N^\mathrm{noise})/N^{\mathrm{total}}.
			    \end{equation}
    
			The number of photoelectrons in each observation obeys the Poisson distribution $\pi(\lambda)$, with the expected number $\lambda$ as
		        \begin{equation}
		            \lambda
		            = -\log(1 - R).
		        \end{equation}
         
\begin{figure}[h]
		            \centering
		            \includegraphics[scale = 0.30]{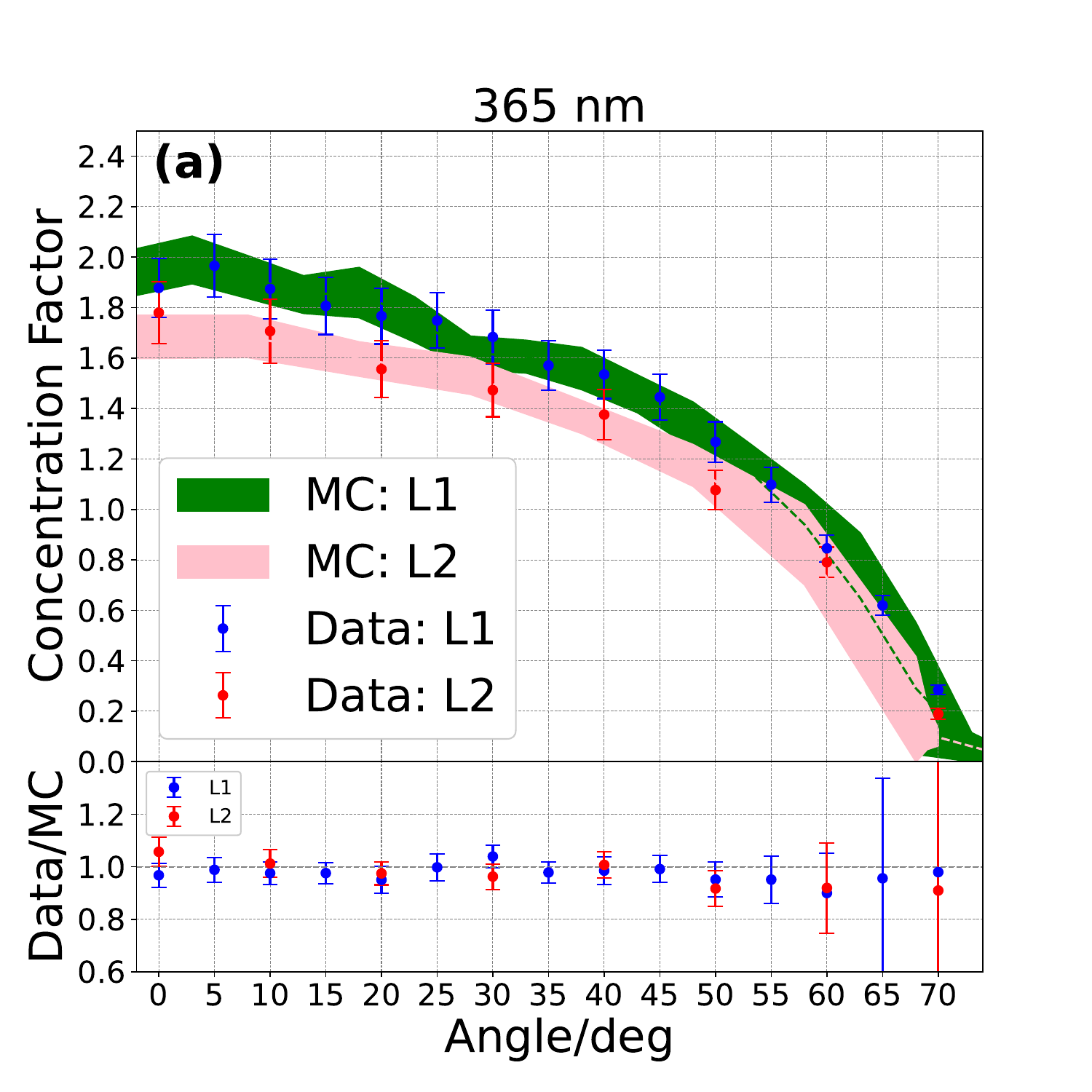}
		            \includegraphics[scale = 0.30]{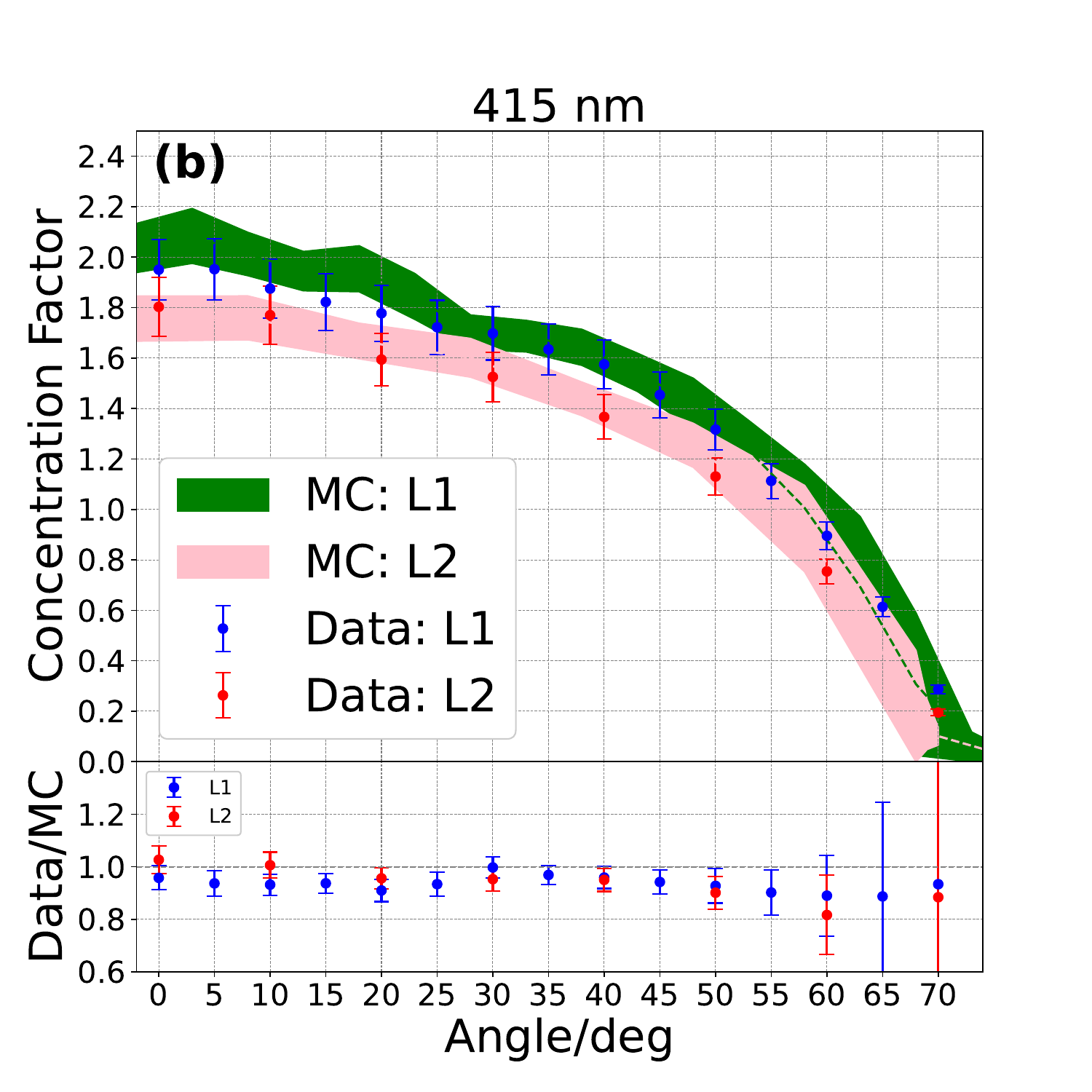}
		            \includegraphics[scale = 0.30]{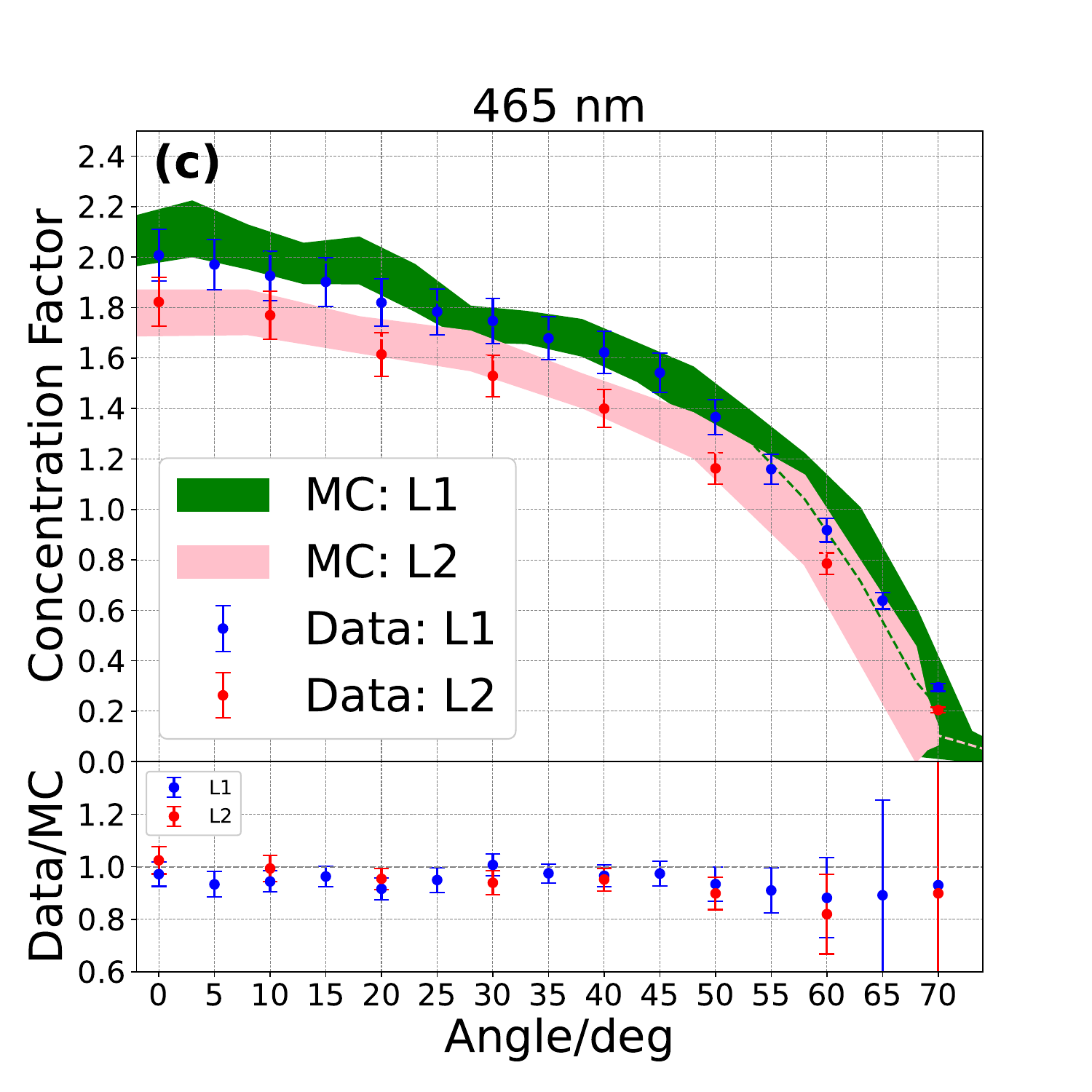}
		            \includegraphics[scale = 0.30]{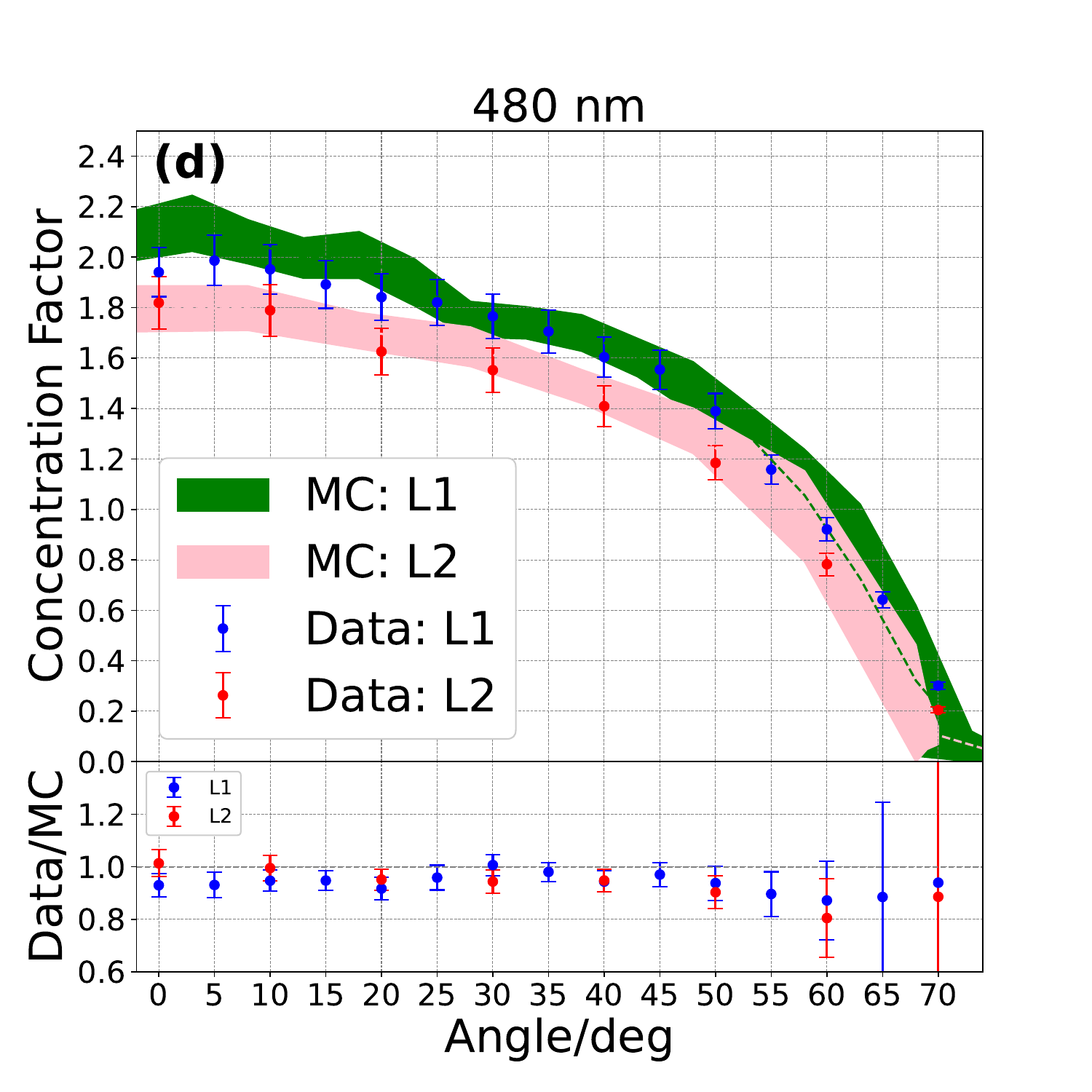}
		            \caption{Angular responses of the concentration factors of data  and MC for four wavelengths using the diffusion ball at rail L1 and L2. Top left (right) represents 365 (415) nm, and bottom left (right) represents 465 (480) nm. The ratio of data/MC panels show that the data and MC are in good agreement. The error bars and bands reflect total uncertainty including statistical and systematic uncertainty.}  
		            \label{fig:concentrationfactor} 
		        \end{figure}

			To eliminate the instability of the light source, we use the PE ratio $\eta \equiv \lambda_\mathrm{test}/\lambda_\mathrm{calib}$, where $\lambda_\mathrm{test}$ and $\lambda_\mathrm{calib}$ are $\lambda$s for the test and calibration PMTs, respectively.  This ratio is independent of the intensity of the light source. The concentration factor $\eta^\mathrm{con}/\eta^\mathrm{bare}$ represents the boost in DE due to the concentrator, where $\eta^\mathrm{con}$ and $\eta^\mathrm{bare}$  are the PE ratio in the cases with and without the concentrator on the test PMT. The points in Fig.~\ref{fig:concentrationfactor} show the measured angular responses of the concentration factors for four wavelengths at rail L1 and L2.  The incident angle is defined as the angle between the vector from the center of the light concentrator to the center of the light source and the normal of the PMT.

	\subsection{Monte Carlo simulations}
		Based on the geometry and physics of the PMT and the associated concentrator described in Sects.~\ref{sssec:design} and \ref{sssec:material}, we carried out a MC simulation to obtain the concentration factor.  The simulation setup is the same as the experimental one, as shown in Fig.~\ref{fig:DarkBox}. The optical photons are isotropically emitted by a spherical source with diameter of $5$ cm, and the measured reflectivity of the coated aluminum (shown in Fig.~\ref{fig:concentratorref}) is used as input for the simulation. For simplicity, we assume the absorptance of the dark box to be unity. The ratio $\eta$ is calculated with the number of photons that hit the test and calibration PMTs and used to estimate the concentrator factor. The comparison of the concentrator factors between the simulation and the experiment can be found in Fig.~\ref{fig:concentrationfactor}. Good agreement is achieved between the simulation and the experiment. 
		
\begin{figure}[h]
                \centering
                \includegraphics[scale = 0.7]{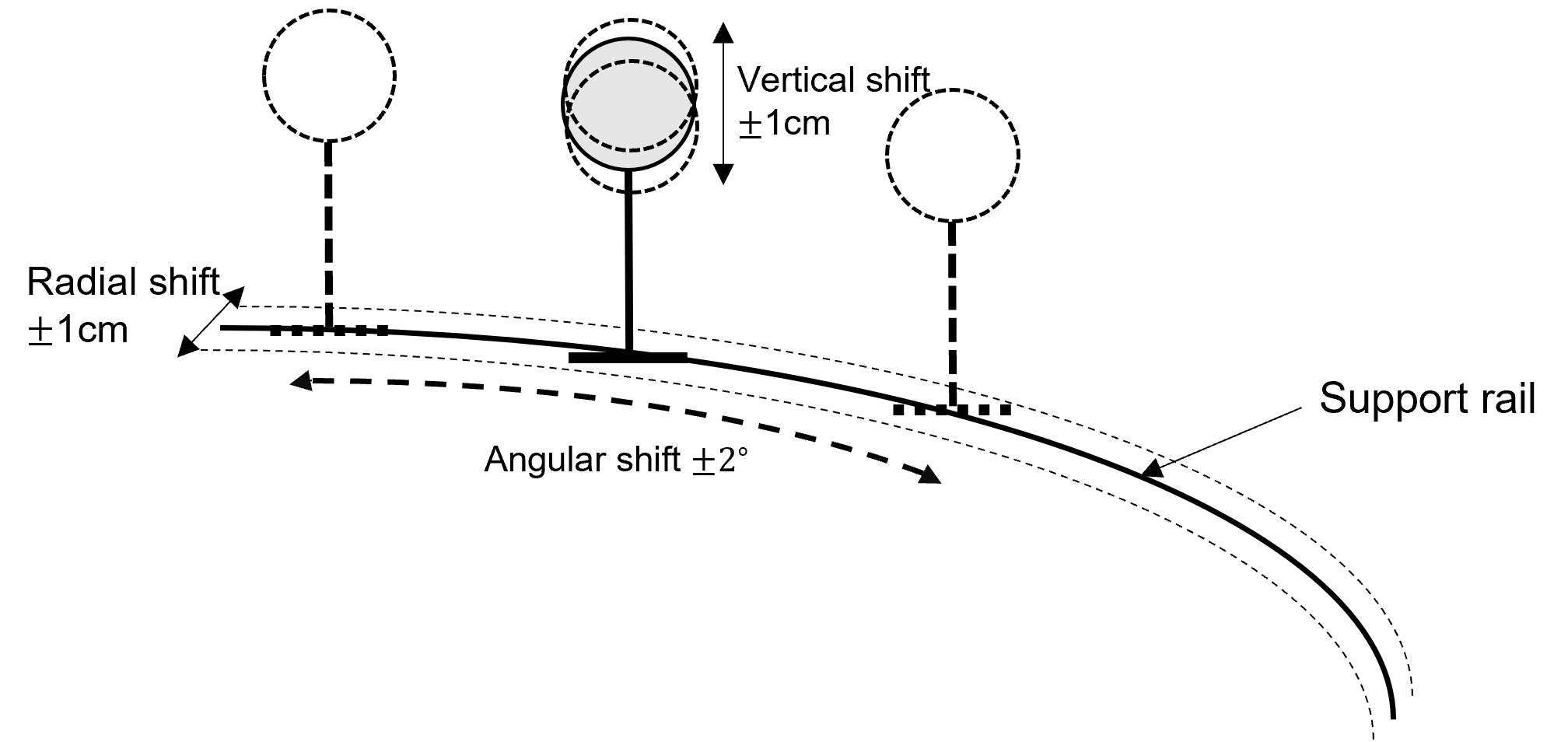}
                \caption{Schematic for angular, vertical and radial shifts.}
                \label{Pic: schematic}
            \end{figure}

		We found that moving the light source around the rail can significantly affect the curve of the concentrator factor, especially if the incident angle is greater than $50 ^{\circ}$. In reality, the manufacturing artifact can bias the PMT’s normal by a few degrees. 
		By matching data points and the MC curve, the angular bias of $3\pm 2^{\circ}$ is found and applied in the estimation of systematic uncertainty, i.e., the MC curve is shifted by $+3 ^{\circ}$ as the central line and further shifted by $\pm 2^{\circ}$ as the systematic uncertainty shown as the band in Fig.~\ref{fig:concentrationfactor}. The systematic uncertainties resulting from vertical and radial shifts of the light source by $\pm 1$ cm are also studied.  In the incident angle range $0-50^{\circ}$, these three systematic errors are similar with a value less than $5\%$, while the angular uncertainty becomes dominant beyond $50^{\circ}$.  The schematic (Fig.~\ref{Pic: schematic}) shows the operation of the angular, vertical, and radial shifts.

	\subsection{Systematic uncertainty}
	    Two systematic uncertainties for the experimental measurement are considered, which are summarized in Table~\ref{Table_Uncertainties}. In the measurement, for each wavelength, we used the diffusion ball of different diameters of $2.0$, $3.0$, $4.0$, and $5.0$ cm, respectively, placing them at a fixed incident angle of $0^{\circ}$. To be conservative, we assign the maximum difference between the mean and the largest/smallest of the measured concentrator factors as the systematic uncertainty, which is less than 6.1\% for the four wavelengths. The second uncertainty arises from the installation of the light concentrator. For each wavelength, we measured the concentrator factor in five installation sets using the case of the $5.0$ cm diffusion ball and the incident angle of $0^{\circ}$ as the base. The method for estimating the installation uncertainty is the same as that for the diameter. 
		\begin{table}
			\caption{Summary of systematic uncertainties for the experimental measurement. The second and third columns list the uncertainty from the diameter of the diffusion ball and the installation of the light concentrator, and the fourth one is the total systematic uncertainty which is the quadratic sum of the second and third columns. }
			\label{Table_Uncertainties}
				\begin{tabular*}{\tblwidth}{@{}CCCC@{}}
					\toprule
					                Wavelength (nm) & Diameter & Installation & Total\\
					\midrule
					                $365$ & $3.8\%$ & $4.9\%$ & $6.2\%$\\
					                $415$ & $6.1\%$ & $1.0\%$ & $6.2\%$\\
					                $465$ & $4.9\%$ & $1.2\%$ & $5.1\%$\\
					                $480$ & $4.9\%$ & $0.8\%$ & $5.0\%$\\
					\bottomrule
				\end{tabular*}
		\end{table}

\section{Discussion}\label{sssec:discussion}
	\subsection{Low background}
		The JNE experiment aims to detect rare neutrino signals in the deep laboratory. To minimize interference from other background radiation as much as possible, we should use low-radioactivity background materials to construct the detector, including the stainless steel, acrylic, ropes, and PMTs. Currently, the mass of one concentrator is approximately 4 kg, about 12 tons of ABS plastic are needed to make all the concentrators. When choosing ABS materials, it is necessary to ensure the high cleanliness of the raw materials and processing procedures to minimize the impact of the radioactive background of ABS plastic on neutrino detection. 

	\subsection{The protection of aluminum film}
		During the trial and testing process of the concentrator sample, the aluminum film used did not have a protective coating. Considering that the concentrator will be stored in air, installed, and used in pure water in the future, it is difficult to guarantee the stability of the naked aluminum film. For instance, oxidation reactions can cause the aluminum film to turn black, reducing its reflectivity. To ensure the long-term stability and optical performance of the aluminum film, we refer to the method of the SNO experiment~\cite{doucas1996light} and propose to apply a protective layer to the aluminum film. Although the introduction of the protective layer may have a certain impact on the reflectivity of the aluminum film, by choosing the appropriate material and designing a reasonable coating thickness, this impact can be minimized.

	\subsection{Improvement of light collection efficiency and energy resolution}
\begin{figure}[h]
		        \centering
		        \includegraphics[scale=0.40]{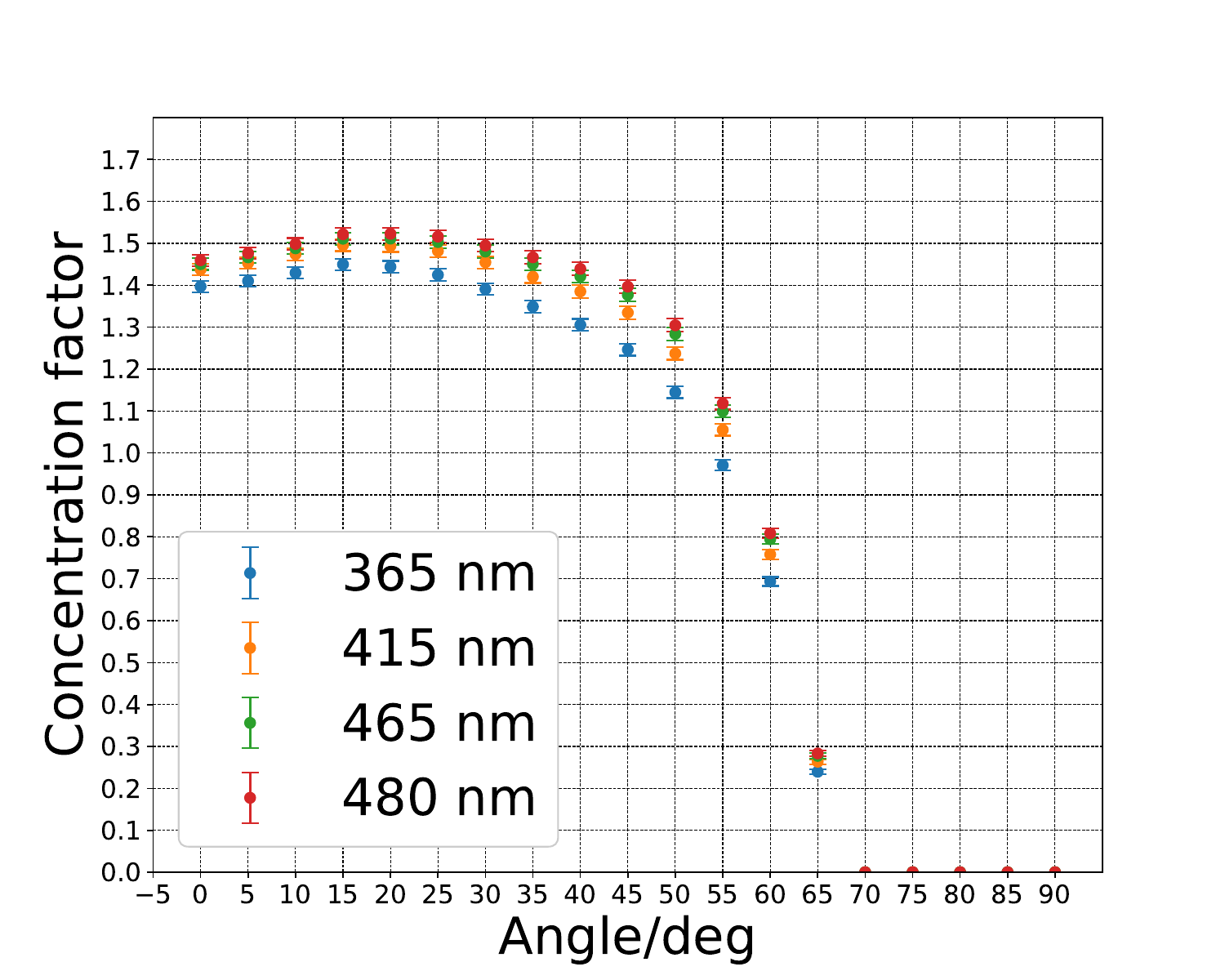}
		        \caption{Angular responses of the concentration factors for four wavelengths using the simulated parallel light. }
		        \label{fig:cfactorparallel}
		    \end{figure}

		For the same wavelength, the concentrator factor curves at different rails are different due to the geometric effect. For a PMT in a large-scale detector, the incident light approximates to parallel for events around the center of the detector. MC simulations using a parallel light source instead of the diffusion ball are performed with other conditions unchanged to eliminate the geometric effect. The angular responses of the concentrator factors, in this case, are shown in Fig.~\ref{fig:cfactorparallel}. Note that here the concentrator factor is defined as the ratio of the number of photons that hit the region of the exit aperture with the concentrator to that without it, i.e. the area below the mask is not accounted for. By this definition, the contribution of a more significant TTS part around the equator of PMT, as shown in the right panel of Fig.~\ref{fig:PMTTTS}, is eliminated. The curve of the concentrator factor moves upward as the wavelength increases due to  the higher reflectivity. In general, the light collection efficiency is improved by approximately $40\%$ for parallel light when the incident angle is less than 30$^\circ$. Correspondingly, the improvement in energy resolution is estimated to be around $15\%$ with the assumption that the energy resolution scales inversely proportional to the square root of the number of detected photons.
    
\section{Conclusion}\label{sssec:conclusion}
	A custom-designed light concentrator is developed to match the NNVT 8-inch MCP-PMT of the JNE. A light bowl with a cut-off angle of $70^{\circ}$ is selected to accommodate the maximum angular tolerance while maintaining a high collection efficiency. Tensile test and finite element analysis confirm that ABS plastic provides sufficient mechanical support for the photon detecting device and is expected to have long-term stability. Experimental measurements of the angular responses of the photon detecting device using a diffusion ball were performed to compare with the Monte Carlo simulation, and the results show good agreement. MC simulations with parallel light show that the light concentrator can improve the efficiency of light collection by $40\%$. With the combination of PMT and light concentrator, it is expected that the energy resolution will improve by more than 15\% for events at the center of the JNE detector, and the TTS will only increase by no more than 0.3 ns (FWHM). These results confirm that this custom-designed light concentrator is well-suitable for the JNE. 

	\subsection* {Acknowledgments}
		This work is supported in part by the Discipline Construction Fund of Shandong University, the Key Laboratory of Particle \& Radiation Imaging (Tsinghua University), and the CAS Center for Excellence in Particle Physics (CCEPP), the National Natural Science Foundation of China (Nos. 12127808, 12141503, and 11620101004), the China Postdoctoral Science Foundation (Certificate Number: 2024M751611), and the Ministry of Science and Technology of China (No. 2022YFA1604704). We thank the State Key Laboratory of New Ceramics \& Fine Processing of Tsinghua University for measuring the reflectivity of coated aluminum.  The authors acknowledge Orrin Science Technology and Jingyifan Co., Ltd for the efforts in the engineering design and fabrication of the dark box. 

\bibliographystyle{unsrt}

\bibliography{cas-refs-new}

\begin{thebibliography}{10}

\bibitem{Jinping:2016iiq}
John~F. Beacom et~al.
\newblock {Physics prospects of the Jinping neutrino experiment}.
\newblock {\em Chin. Phys. C}, 41(2):023002, 2017.

\bibitem{Ma:2021uzi}
Hao Ma, Wenhan Dai, Zhi Zeng, Tao Xue, Litao Yang, Qian Yue, and Jianping
  Cheng.
\newblock {Status and prospect of China Jinping Underground Laboratory}.
\newblock {\em J. Phys. Conf. Ser.}, 2156(1):012170, 2021.

\bibitem{Super-Kamiokande:2002weg}
Y.~Fukuda et~al.
\newblock {The Super-Kamiokande detector}.
\newblock {\em Nucl. Instrum. Meth. A}, 501:418--462, 2003.

\bibitem{SNO:1999crp}
J.~Boger et~al.
\newblock {The Sudbury neutrino observatory}.
\newblock {\em Nucl. Instrum. Meth. A}, 449:172--207, 2000.

\bibitem{DayaBay:2012fng}
F.~P. An et~al.
\newblock {Observation of electron-antineutrino disappearance at Daya Bay}.
\newblock {\em Phys. Rev. Lett.}, 108:171803, 2012.

\bibitem{Borexino:2008gab}
G.~Alimonti et~al.
\newblock {The Borexino detector at the Laboratori Nazionali del Gran Sasso}.
\newblock {\em Nucl. Instrum. Meth. A}, 600:568--593, 2009.

\bibitem{IceCube:2016zyt}
M.~G. Aartsen et~al.
\newblock {The IceCube Neutrino Observatory: Instrumentation and Online
  Systems}.
\newblock {\em JINST}, 12(03):P03012, 2017.
\newblock [Erratum: JINST 19, E05001 (2024)].

\bibitem{adrian2016letter}
Silvia Adrian-Martinez, M~Ageron, F~Aharonian, S~Aiello, A~Albert, F~Ameli,
  E~Anassontzis, M~Andre, G~Androulakis, M~Anghinolfi, et~al.
\newblock Letter of intent for km3net 2.0.
\newblock {\em Journal of Physics G: Nuclear and Particle Physics},
  43(8):084001, 2016.

\bibitem{Luo:2023reconstruction}
Wentai Luo, Qian Liu, Yangheng Zheng, Zhe Wang, and Shaomin Chen.
\newblock {Reconstruction algorithm for a novel Cherenkov scintillation
  detector}.
\newblock {\em Journal of Instrumentation}, 18(02):P02004, 2023.

\bibitem{wu2023performance}
Yiyang Wu, Jinjing Li, Shaomin Chen, Wei Dou, Lei Guo, Ziyi Guo, Ghulam
  Hussain, Ye~Liang, Qian Liu, Guang Luo, et~al.
\newblock {Performance of the 1-ton prototype neutrino detector at CJPL-I}.
\newblock {\em Nuclear Instruments and Methods in Physics Research Section A:
  Accelerators, Spectrometers, Detectors and Associated Equipment},
  1054:168400, 2023.

\bibitem{Xu:2022wcq}
Xun-Jie Xu, Zhe Wang, and Shaomin Chen.
\newblock {Solar neutrino physics}.
\newblock {\em Prog. Part. Nucl. Phys.}, 131:104043, 2023.

\bibitem{shao2023potential}
Wenhui Shao, Weiran Xu, Ye~Liang, Wentai Luo, Tong Xu, Ming Qi, Jialiang Zhang,
  Benda Xu, Zhe Wang, and Shaomin Chen.
\newblock The potential to probe solar neutrino physics with licl water
  solution.
\newblock {\em The European Physical Journal C}, 83(9):799, 2023.

\bibitem{Bellini:2013wsa}
G.~Bellini, A.~Ianni, L.~Ludhova, F.~Mantovani, and W.~F. McDonough.
\newblock {Geo-neutrinos}.
\newblock {\em Prog. Part. Nucl. Phys.}, 73:1--34, 2013.

\bibitem{wan2017geoneutrinos}
Linyan Wan, Ghulam Hussain, Zhe Wang, and Shaomin Chen.
\newblock Geoneutrinos at jinping: Flux prediction and oscillation analysis.
\newblock {\em Physical Review D}, 95(5):053001, 2017.

\bibitem{wang2020hunting}
Zhe Wang and Shaomin Chen.
\newblock Hunting potassium geoneutrinos with liquid scintillator cherenkov
  neutrino detectors.
\newblock {\em Chinese Physics C}, 44(3):033001, 2020.

\bibitem{DeGouvea:2020ang}
Andr\'e De~Gouv\^ea, Ivan Martinez-Soler, Yuber~F. Perez-Gonzalez, and
  Manibrata Sen.
\newblock {Fundamental physics with the diffuse supernova background
  neutrinos}.
\newblock {\em Phys. Rev. D}, 102:123012, 2020.

\bibitem{wei2017discovery}
Hanyu Wei, Zhe Wang, and Shaomin Chen.
\newblock Discovery potential for supernova relic neutrinos with slow liquid
  scintillator detectors.
\newblock {\em Physics Letters B}, 769:255--261, 2017.

\bibitem{Dolinski:2019nrj}
Michelle~J. Dolinski, Alan W.~P. Poon, and Werner Rodejohann.
\newblock {Neutrinoless Double-Beta Decay: Status and Prospects}.
\newblock {\em Ann. Rev. Nucl. Part. Sci.}, 69:219--251, 2019.

\bibitem{fu2024comparison}
Hao-Yang Fu, Wen-Tai Luo, Xiang-Pan Ji, and Shao-Min Chen.
\newblock Comparison study of counting and fitting methods in search for
  neutrinoless double beta decays.
\newblock {\em arXiv:2412.19859}, 2024.

\bibitem{Wang:2024upf}
Zongyi Wang, Yuhao Liu, Shaomin Chen, Yuanqing Wang, Zhe Wang, and Ming Huang.
\newblock {Structural design of the acrylic vessel for the Jinping Neutrino
  Experiment}.
\newblock {\em JINST}, 19(07):P07041, 2024.

\bibitem{Guo:2017nnr}
Ziyi Guo, Minfang Yeh, Rui Zhang, De-Wen Cao, Ming Qi, Zhe Wang, and Shaomin
  Chen.
\newblock {Slow Liquid Scintillator Candidates for MeV-scale Neutrino
  Experiments}.
\newblock {\em Astropart. Phys.}, 109:33--40, 2019.

\bibitem{Shao:2022yjc}
Wenhui Shao, Weiran Xu, Ye~Liang, Wentai Luo, Tong Xu, Ming Qi, Jialiang Zhang,
  Benda Xu, Zhe Wang, and Shaomin Chen.
\newblock {The potential to probe solar neutrino physics with LiCl water
  solution}.
\newblock {\em Eur. Phys. J. C}, 83(9):799, 2023.

\bibitem{Zhang:2023ued}
Aiqiang Zhang, Benda Xu, Jun Weng, Huiyou Chen, Wenhui Shao, Tong Xu, Ling Ren,
  Sen Qian, Zhe Wang, and Shaomin Chen.
\newblock {Performance evaluation of the 8-inch MCP-PMT for Jinping Neutrino
  Experiment}.
\newblock {\em Nucl. Instrum. Meth. A}, 1055:168506, 2023.

\bibitem{Lay:1996tp}
M.~D. Lay and M.~J. Lyon.
\newblock {An experimental and Monte Carlo investigation of the R1408 Hamamatsu
  8-inch photomultiplier tube and associated concentrator to be used in the
  Sudbury Neutrino Observatory}.
\newblock {\em Nucl. Instrum. Meth. A}, 383:495--505, 1996.

\bibitem{SNO:2022qvw}
A.~Allega et~al.
\newblock {Evidence of Antineutrinos from Distant Reactors using Pure Water at
  SNO+}.
\newblock {\em Phys. Rev. Lett.}, 130(9):091801, 2023.

\bibitem{Oberauer:2003ac}
L.~Oberauer, C.~Grieb, F.~von Feilitzsch, and I.~Manno.
\newblock {Light concentrators for Borexino and CTF}.
\newblock {\em Nucl. Instrum. Meth. A}, 530:453--462, 2004.

\bibitem{bernlohr2003optical}
K~Bernl{\"o}hr, O~Carrol, R~Cornils, S~Elfahem, P~Espigat, S~Gillessen,
  G~Heinzelmann, G~Hermann, Werner Hofmann, D~Horns, et~al.
\newblock {The optical system of the HESS imaging atmospheric Cherenkov
  telescopes. Part I: layout and components of the system}.
\newblock {\em Astroparticle Physics}, 20(2):111--128, 2003.

\bibitem{radu2000design}
Aurelian~A Radu, John~R Mattox, and Steven Ahlen.
\newblock Design studies for nonimaging light concentrators to be used in very
  high-energy gamma-ray astronomy.
\newblock {\em Nuclear Instruments and Methods in Physics Research Section A:
  Accelerators, Spectrometers, Detectors and Associated Equipment},
  446(3):497--505, 2000.

\bibitem{Loo:2023kij}
Kai Loo.
\newblock {Upgrade of OSIRIS for Future Liquid Scintillator Studies}.
\newblock {\em PoS}, TAUP2023:319, 2024.

\bibitem{shirai2017results}
Junpei Shirai, KamLAND-Zen Collaboration, et~al.
\newblock Results and future plans for the kamland-zen experiment.
\newblock In {\em Journal of Physics: Conference Series}, volume 888, page
  012031. IOP Publishing, 2017.

\bibitem{Aguilar:2014oba}
J.~A. Aguilar, A.~Basili, V.~Boccone, F.~Cadoux, A.~Christov, D.~della Volpe,
  T.~Montaruli, \L{}. P\l{}atos, and M.~Rameez.
\newblock {Design, optimization and characterization of the light concentrators
  of the single-mirror small size telescopes of the Cherenkov Telescope Array}.
\newblock {\em Astropart. Phys.}, 60:32--40, 2015.

\bibitem{Krizan:2024ydm}
Peter Krizan.
\newblock {Novel photon detectors}.
\newblock {\em Nucl. Instrum. Meth. A}, 1065:169482, 2024.

\bibitem{Zhi:2017tbk}
Yu~Zhi, Ye~Liang, Zhe Wang, and Shaomin Chen.
\newblock {Wide field-of-view and high-efficiency light concentrator}.
\newblock {\em Nucl. Instrum. Meth. A}, 885:114--118, 2018.

\bibitem{Weng:2024tjs}
Jun Weng, Aiqiang Zhang, Qi~Wu, Lishuang Ma, Benda Xu, Sen Qian, Zhe Wang, and
  Shaomin Chen.
\newblock {Single electron charge spectra of 8-inch high-collection-efficiency
  MCP-PMTs}.
\newblock {\em Nucl. Instrum. Meth. A}, 1066:169626, 2024.

\bibitem{astm2014standard}
{ASTM International}.
\newblock {\em {Standard test method for tensile properties of plastics}}.
\newblock 2014.

\bibitem{altair2017hyperworks}
{Altair Engineering Inc}.
\newblock Available online: \url{https://altair.com/altair-hyperworks}.

\bibitem{Hibbitt:1984abaqus}
H~D Hibbitt.
\newblock {ABAQUS/EPGEN—A general purpose finite element code with emphasis
  on nonlinear applications}.
\newblock {\em Nuclear Engineering and Design}, 77(3):271--297, 1984.

\bibitem{Yang:2024qqe}
Haoyan Yang, Tao Xue, Lin Jiang, Qiutong Pan, Bo~Liang, Yinong Liu, and Jianmin
  Li.
\newblock {Quantitative analysis of the high speed and high precision waveform
  digital system for Jinping Neutrino Experiment}.
\newblock {\em JINST}, 19(08):T08005, 2024.

\bibitem{xue2018design}
Tao Xue, Jinfu Zhu, Guanghua Gong, Liangjun Wei, Yang Luo, and Jianmin Li.
\newblock The design and data-throughput performance of readout module based on
  zynq soc.
\newblock {\em IEEE Transactions on Nuclear Science}, 65(5):1169--1179, 2018.

\bibitem{doucas1996light}
G~Doucas, S~Gil, NA~Jelley, L~McGarry, ME~Moorhead, NW~Tanner, and CE~Waltham.
\newblock {Light concentrators for the Sudbury Neutrino Observatory}.
\newblock {\em Nuclear Instruments and Methods in Physics Research Section A:
  Accelerators, Spectrometers, Detectors and Associated Equipment},
  370(2-3):579--596, 1996.

\end{thebibliography}

\end{document}